\documentstyle[11pt,aaspp4]{article}
\begin{document}
\title{The Radio Properties of Optically Selected Quasars. III.
Comparison between Optical and X-Ray Selected Samples
\footnote{
Some of the observations reported here
were obtained with the Multiple Mirror Telescope, a facility operated
jointly by the University of Arizona and the Smithsonian
Institution.}}
\author{Eric J. Hooper, Chris D. Impey}
\affil{Steward Observatory, University of Arizona, Tucson, AZ 85721 \\ 
ehooper@as.arizona.edu, cimpey@as.arizona.edu}
\author{Craig B. Foltz}
\affil{Multiple Mirror Telescope Observatory, University of Arizona, 
Tucson, AZ 85721 \\ cfoltz@as.arizona.edu}
\and
\author{Paul C. Hewett}
\affil{Institute of Astronomy, Madingley Road, CB3 0HA Cambridge,
United Kingdom \\ phewett@mail.ast.cam.ac.uk}
\begin{abstract}

A sample of 103 quasars from the Large Bright Quasar Survey (LBQS) has been
observed with the VLA at 8.4 GHz to study the evolution of the radio luminosity
distribution and its dependence on absolute magnitude.  Radio data from pointed
observations are now available for 359 of the 1055 LBQS quasars.  The radio-loud
fraction is constant at $\approx 10\%$ over the absolute magnitude range $-28
\lesssim M_B \lesssim -23$, and it rises to $\sim 20\%$ ($\log R_{8.4} > 1$) or
$\sim 35\%$ ($\log L_{8.4} > 25$) at the brightest absolute magnitudes in the
sample.  This nearly flat distribution differs markedly from those of the
optically selected Palomar-Green (PG) Bright Quasar Survey and the X-ray
selected Extended Medium Sensitivity Survey (EMSS), both of which have lower
radio-loud fractions for absolute magnitudes fainter than $M_B = -24$ and higher
fractions at brighter magnitudes.  The reason for the high radio-loud fraction
at bright absolute magnitudes in the PG, compared to the LBQS and other
optically selected quasar surveys, is unknown.  The trend of increasing
radio-loud fraction with absolute magnitude in the EMSS is due at least in part
to a correlation between X-ray and radio luminosity.  Combining the LBQS data
with radio studies of high-redshift quasars leads to the conclusion that the
radio-loud fraction in optically selected quasars does not appear to evolve
significantly, aside from a modest increase at $z \sim 1$, from $z = 0.2$ to
redshifts approaching 5, a result that is contrary to previous studies which
found a decrease in radio-loud fraction with increasing redshift by comparing
the low-$z$ fraction in the PG to higher redshift samples.

\end{abstract}

\begin{center}
\section{Introduction}
\label{sec:introduction}
\end{center}

Optical and near-UV light from quasars is comprised primarily of
thermal components and emission lines, with some contribution from
synchrotron radiation.  The source of the optical continuum and broad
lines in the standard black hole model for quasars (e.g., Rees
\markcite{Rees84} 1984) is an accretion disk and centrally
concentrated hot gas clouds, respectively.  In this model, optical
luminosity is closely tied to the accretion rate and the mass of the
central engine.

Radiation at radio frequencies is predominantly due to synchrotron
processes, which depend on the energy distribution of relativistic
particles and magnetic field strength.  The ultimate power source for
particle acceleration and magnetic field production is presumably the
same gravitational engine producing the optical flux, but the
relationship is not straightforward.  Radio and optical data for an
ensemble of quasars should place constraints on the connections
between the fueling of the black hole, the gas and dust in the
vicinity of the central engine, and the particle acceleration
mechanism.

The relationship between quasar radio and optical emission was studied
initially using radio-selected objects, which generally had high radio
luminosities due to the relatively low sensitivity limits of the
surveys.  Sandage \markcite{Sandage65} (1965) determined that not
all quasars are powerful radio sources and hence detectable in radio
surveys.  Since then, in addition to radio surveys, radio follow-up
observations have been made of surveys conducted in the optical (e.g.,
Sramek \& Weedman \markcite{SW80} 1980; Condon \markcite{Condon81} et
al. 1981; Marshall \markcite{Marshall87} 1987; Miller, Peacock,
\& Mead \markcite{MPM} 1990; Schmidt et al.\markcite{Schmidt95} 1995)
and other wave bands.  Targeted radio observations of quasars selected
by other means typically go deeper than the radio surveys.  As a
result the median radio luminosity of these samples is lower, but they
usually contain few of the very powerful objects common in radio
surveys.  Taken together the two survey methods have detected quasars
with a range of more than 6 orders of magnitude in radio luminosity.
This wide span of radio power has commonly been divided into two
regimes, radio-loud and radio-quiet.

Two definitions of radio-loud are generally employed, one based on
radio luminosity, and the other on the ratio, $R_{\nu}$, of radio
luminosity at frequency $\nu$ to optical or near-ultraviolet
luminosity.  These different definitions produce the same
classification for most quasars, except at low and high optical
luminosities.  The empirical basis for this was discussed extensively
in Stocke et al.  (1992) and in Hooper et al.\markcite{PaperII} (1995;
hereafter Paper II).  Peacock, Miller, \& Longair \markcite{Peacock86}
(1986) explored the implications of using radio luminosity or a
radio-to-optical luminosity ratio as the parameter of radio strength
in a joint radio-optical luminosity function.  The analysis in the
present work is conducted using both definitions.

Many studies have concluded that the fraction of radio-loud quasars
decreases with increasing redshifts by comparing the radio
observations of the predominantly low-redshift Palomar-Green optical
quasar survey (Kellermann et al.\markcite{Kellermann89} 1989,
hereafter PG), which has a high radio-loud fraction, to higher
redshift optically selected samples with relatively fewer radio-loud
quasars (Peacock et al.\markcite{Peacock86} 1986; Miller et
al.\markcite{MPM} 1990; Schneider et al.\markcite{Schneider92} 1992;
Visnovsky et al.\markcite{Visnovsky92} 1992).  However, there is no
strong evidence for evolution for $z > 1$ if the PG is excluded from
the ensemble of optically selected samples (La Franca et
al.\markcite{LaF94} 1994).  A sample of quasars from the Large Bright
Quasar Survey (LBQS; Hewett, Foltz, \& Chaffee 1995) in the redshift
range $0.2 < z < 3.4$ did not show the strong evolution in radio-loud
fraction inferred from comparisons of the PG with other surveys (Paper
II\markcite{PaperII}).  These results indicate that the PG has an
anomalously high radio-loud fraction (at bright absolute magnitudes)
compared to other optically selected quasar samples.  This difference
remains unexplained.

Peacock et al. (1986) noted that strong radio emitters were found to
be relatively rare at absolute blue magnitudes ($M_B$) fainter
than $-24$.  These authors proposed a simple and plausible selection
effect, arising from the classification of active galactic nuclei
(AGN) as either quasars or radio galaxies, to explain this
observation.  If radio-loud AGN reside in large bright elliptical
galaxies, and radio-quiet AGN have less luminous spiral host galaxies
(see, e.g., Smith et al. \markcite{Smith86} 1986; V\'{e}ron-Cetty \&
Woltjer \markcite{VCW90} 1990; Hutchings et al.\markcite{Hutchings94};
however, there are indications that some radio-quiet quasars reside in
elliptical host galaxies: e.g., Disney et al.\markcite{Disney95} 1995;
Bahcall, Kirhakos, \& Schneider\markcite{Bahcall96} 1996), powerful
radio sources with optically weak nuclei would be classified as radio
galaxies in radio surveys.  They would not be selected by optical
surveys because of the low contrast between the quasar light and the
relatively bright, red, and extended stellar component.  In fact, just
such a decrease in the fraction of radio-loud quasars with $M_B >
-24$ has been seen directly in follow-up radio data to optical and
X-ray surveys.  The effect is highly significant in the PG survey (see
also Padovani\markcite{Padovani93} 1993; La Franca et
al.\markcite{LaF94} 1994), and a sharp decrease in radio-loud fraction
at the same optical luminosity is present among the quasars selected
at X-ray wavelengths by the Extended Medium Sensitivity Survey (EMSS;
Della Ceca et al.\markcite{Della94} 1994).  In apparent support of
this conclusion, no radio-loud quasars fainter than $M_B = -24$ were
found in a sample from the LBQS discussed in Paper
II\markcite{PaperII}.  The probability of selecting a quasar using the
LBQS criteria is quantifiable in terms of quasar spectral energy
distribution, absolute magnitude, redshift, and brightness relative to
the host galaxy, an advantage over many existing surveys.  This
probability distribution was used to test the selection effect
hypothesis of Peacock et al. (1986), which was found to not explain
the trend seen in the sample drawn from the LBQS.

Further analysis of the radio properties of optically selected quasars
with absolute magnitudes near $M_{B} = -24$ was the primary motivation
for obtaining the data presented in the current work, which is the
third paper in a series examining the radio properties of the LBQS.
Paper I in the series (Visnovsky et al.\markcite{Visnovsky92} 1992)
presented radio data for 124 high redshift ($1.0 < z < 3.0$) LBQS
quasars, selected as a comparison sample for the predominantly low
redshift PG survey.  The parameter $R_{8.4}$ was found to be
independent of both redshift and absolute magnitude in this subsample,
and the distribution of $\log R_{8.4}$ appeared to be bimodal.  The
radio-loud fraction was lower than in the PG sample, from which it was
inferred that the fraction evolves.  One of the principal motivations
at that point for expanding the LBQS radio sample was to cover the
full redshift range of the LBQS ($0.2 < z < 3.4$), in order to
independently investigate the evolution of the radio-loud fraction
using a single well-defined survey.  An additional 132 LBQS quasars
were added to the radio sample in Paper II to accomplish this.  The
optically brightest quasars at each redshift were selected for Papers
I and II in order to study the radio properties at the lowest values
of $R_{8.4}$ possible for a given radio flux limit.  A more detailed
discussion of sample selection is given in Section 2.1.  The
radio-loud fraction in the Paper II sample was nearly constant at
$\sim 10\%$ from redshift 0.2 to 3.4, and, with the addition of
high-$z$ samples from the literature, showed no evolution to a
redshift of almost 5.  The fraction was also independent of absolute
blue magnitude in the range $-27.5 < M_{B} < -24$.  A higher
radio-loud fraction was found at the brightest absolute magnitudes.
As discussed above, there were no radio-loud quasars with $M_{B} >
-24$.

The change in radio-loud fraction of LBQS quasars at faint absolute
magnitudes reported in Paper II\markcite{PaperII}, while significant
at the 97\% confidence level, was based on only 20 quasars with $M_B $
fainter than $-24$.  Absolute magnitude errors were $0.3$ to $0.4$ for
$M_B \sim -24$.  Given the survey's relative insensitivity to
selection bias against optically weak, radio-loud quasars (see Paper
II and Hewett et al.\markcite{Hewett95} 1995 for details), it is
important to expand the radio database of LBQS quasars and to reduce
the magnitude uncertainties of radio-loud objects with absolute
magnitudes near $M_B = -24$.  New observations with the Very Large
Array\footnote{The Very Large Array (VLA) of the National Radio
Astronomy Observatory is operated by Associated Universities, Inc.,
under a cooperative agreement with the National Science Foundation.}
(VLA) of an additional 103 LBQS quasars with absolute magnitudes in
the range $-25 < M_B < -23$ and redshifts $0.2 < z < 1.1$ are
presented in this paper, along with improved magnitude determinations
for objects at faint optical $M_B$ (Section \ref{sec:data}).  The
latest data are combined with those from Papers
I\markcite{Visnovsky92} and II\markcite{PaperII} to form one of the
largest samples of sensitive radio observations of quasars selected at
other wavelengths in a single survey.  The distribution of radio
luminosity as a function of $M_B$ is analyzed and compared to the PG
and EMSS samples (Section \ref{sec:MB}), and the evolution of the
radio properties of these three samples is explored in Section
\ref{sec:z}.  Models proposed to explain the radio luminosity
distribution as a function of absolute magnitude are discussed in
Section \ref{sec:discussion}.  Section \ref{sec:summary} contains a
summary of the major results.

\pagebreak 

\section{Observations and Data Reduction}
\label{sec:data}

Radio observations of an additional sample of LBQS quasars and
improved optical data for a subset of the radio-loud quasars are
presented and discussed below.  Rest-frame 8.4 GHz luminosities ($L_{8.4}
\!$) and absolute blue magnitudes ($M_B$) were calculated using
$H_{0} = 50$ km ${\rm s^{-1}}$ ${\rm Mpc^{-1}}$ and $q_{0} = 0.5$.
The luminosity ratio, $R_{8.4} \!$, is $L_{8.4} $ divided by the average
luminosity over the $B$ passband.

The radio luminosities were derived from observed fluxes assuming a
radio spectral index $\alpha = -0.5$ $(f_{\nu } \propto \nu ^{\alpha
})$.  Optical $k$-corrections for the majority of the quasars in the
sample were calculated from the composite quasar spectrum of Francis
et al.\markcite{Francisetal91} (1991).  Details of these calculations
and attendant error analysis are given in Paper II\markcite{PaperII}.
The absolute magnitude errors, which are conservative overestimates,
are typically $\pm 0.4$ mag for $M_B \approx -24$, rising at brighter
$M_B$ (higher redshifts) to $\pm 0.9$.   These errors are the
quadrature sum of three components: 1) uncertainty in the
$k$-correction, the dominant source of error for most of the quasars
in the sample, particularly at high redshifts; 2) optical variability
of the quasars over the time interval between the optical and radio
observations; and 3) a photometric imprecision of $\approx 0.15$ mag
in the original $B_{J}$ magnitudes determined from Schmidt photographic
plates (Hewett et al.\markcite{Hewett95} 1995), which is the smallest
contributor to the total uncertainty, although still significant at
faint absolute magnitudes.  While the resulting precision is adequate
for determining large-scale trends in a substantial data set, a
detailed view of the change in radio-loud fraction around $M_B = -24$
is difficult with the current magnitude uncertainties, given the small
number of radio-loud quasars.  Each of these uncertainties was reduced
for several radio-loud quasars with absolute magnitudes near $M_B =
-24$ by obtaining improved optical spectra and multicolor CCD
photometry at an epoch closer to that of the radio data.

\subsection{Sample Selection}
\label{ssec:sample_selection}

Each subsample of the LBQS selected for observing with the VLA was
chosen to address specific scientific questions, as discussed in the
introduction.  Neither the individual subsamples nor the combined set
of radio data are representative of the LBQS as a whole, in the sense
that the distributions of $B$, $z$, and $M_{B}$ are different for the
radio sample and the remaining LBQS quasars (see Figure 1 in Paper
II\markcite{PaperII}).  The current sample is simply that from Paper
II\markcite{PaperII} with the addition of many of the remaining LBQS
quasars in the magnitude range $-25 < M_{B} < -23$.  The radio sample
now contains 146/276 (53\%) of all LBQS quasars in this absolute
magnitude range.  The 103 new observations are nearly equally divided
about $M_{B} = -24$.  Figure \ref{fig:zMBLBQS} demonstrates the
effects of the various selection criteria on the radio sample.
Absolute blue magnitude is plotted against redshift, with individual
radio detections and upper limits indicated by filled circles and
horizontal lines, respectively, and the Y-shaped symbols represent
LBQS quasars without radio data.  All of the quasars without radio
data are plotted for $z > 2.5$, but, due to the large numbers of such
objects at lower redshifts, only 1/3 of them with $z < 2.5$ are
plotted.

Figure \ref{fig:zMBLBQS} shows that, for $M_{B} < -25$, the radio
sample lies in the brighter portion of the region of absolute
magnitude/redshift space occupied by the LBQS.  This effect is due to
the selection priorities of Papers I\markcite{Visnovsky92} and
II\markcite{PaperII}, which resulted in the exclusion of almost all
LBQS quasars with $B > 18.4$ from those samples.  There is a thin
strip corresponding to $B > 18.4$ in which there are no radio data,
and the radio properties of the LBQS are formally unconstrained for
these combinations of $z$ and $M_{B}$.  However, it is unlikely that
the radio properties are radically different in this unsampled region.
The strip is narrow, typically half a magnitude at any redshift.  The
difference between the brightest members of the radio sample and the
remaining LBQS quasars was measured in a series of adjoining redshift
intervals of width $0.2$ over the range $1.1 < z < 3.3$.  The average
magnitude difference is $0.55$ mag, with an rms dispersion about the
mean of $0.15$ mag.  The unsampled region spans between 125 Myr and 1
Gyr of lookback time.  Therefore, any major differences between the
sample and the unsampled region would have to occur over a luminosity
interval of 0.5 mag and $\leq 1$ Gyr in lookback time.  These are
small intervals for major changes, given that the radio properties of
LBQS quasars for which data exist are nearly constant over 5
magnitudes and 8 Gyr of lookback time (see Sections \ref{sec:MB} and
\ref{sec:z}).  Note that, for every value of redshift and absolute
magnitude in the unsampled strip, there are radio data at a different
$z$/$M_{B}$.

The unsampled region extended over all absolute magnitudes in Paper
II\markcite{PaperII}, but it ends at $M_{B} = -25$ in the present work
due to the addition of the new radio data in the range $-25 < M_{B} <
-23$.  The radio sample now covers the same region of absolute
magnitude/redshift space as the entire LBQS in this absolute magnitude
range.  No radio data are available for LBQS quasars fainter than
$M_{B} = -22.7$ (Figure \ref{fig:zMBLBQS}), but this does not affect
the conclusions regarding the radio properties of the brighter quasars
in the sample.

\subsection{Radio}
\label{ssec:dataradio}

All of the radio observations of LBQS quasars were obtained with the
VLA at a frequency of 8.4 GHz.  The latest group of 103 quasars,
presented here, were observed on UT 1993 June 4 and 25 in the B/C and
C array configurations, respectively.  Exposure times were about 2
minutes for all objects.  Standard AIPS software was employed to
produce dirty total-intensity (Stokes I) maps for all fields.  Those
containing a strong source were subsequently CLEANed.

A detailed description of the noise analysis and source detection
procedure used for these maps was presented in Paper II, but the
principal steps are summarized here.  Radio flux at any point on a map
was determined by fitting the core of the corresponding beam to the
region around the desired map location.  Noise levels were determined
by performing the fitting procedure on a random sample of points
across each map.  Subsequently, a circle of radius $2''$ at the center
of the map was searched for radio flux exceeding a detection threshold
of three times the rms noise value.

A small number of spurious detections are expected near the $3 \sigma$
detection limit.  The probability of obtaining a $3 \sigma$ or greater
positive fluctuation at a given position on an apparently empty map in
the latest set of observations is $1.9 \times 10^{-3}$, 1.4 times
greater than predicted by a Gaussian noise model using the measured
rms.  The excess of positive fluctuations may be due to weak sources
in the maps, or it may be due to part of the sidelobe patterns from
unknown off-map sources.  The chance of a spurious detection in a map
is this probability multiplied by the number of independent sampling
points within the $2''$ search radius.  The average number of
independent sample points per search disk for the current dataset is
3.4, giving an expected total of $(1.9 \times 10^{-3})(3.4)103 \approx
0.7$ spurious detections among the 103 maps.  The expected number of
spurious detections in the previous radio data sets is 2.4 (Paper
II\markcite{PaperII}), for a total of 3.1 among 61 recorded detections
from 359 observed fields.

The search radius was chosen to match the astrometric accuracy of the
LBQS, estimated to be $\lesssim 2''$ for the original quasar
positions.  The results were checked by reanalyzing the data with a
$4\farcs5$ search radius to look for larger positional offsets.  Every
strong source (flux $ > 7 \sigma$, where $\sigma $ is the rms noise)
found using the larger search radius is within $2''$ of the map
center, and the greatest distance from the center of any detected
source is $2\farcs4$ (1322$-$0204, a $3.8 \sigma $ detection).  One
quasar not detected using a $2''$ radius had a $> 3\sigma $ positive
fluctuation within a $4\farcs5$ radius.  The expected number of
additional spurious detections using the larger radius in 103 maps is
$\sim 2.7$.  These tests indicate that a $2''$ search radius is
adequate, and this value was used to determine the LBQS radio fluxes.

Radio and optical data for each quasar in the current data set, as
well as quasars listed in Paper II\markcite{PaperII} with new optical
data, are listed in Table \ref{tab:tablerad} in the following format:

\begin{itemize}

\item[ ]
Column (1):  Object name, listed in order of increasing right
ascension.  

\item[ ]
Column (2):  Apparent $B$-magnitude derived from observed $B_{J}$ (Hewett
et al.\markcite{Hewett95} 1995) as described in Paper II, or, for 5
quasars flagged in the table, measured directly (Section
\ref{ssec:dataoptical}).  

\item[ ]
Column (3):  Redshift from Hewett et al.\markcite{Hewett95} (1995).
 
\item[ ]
Column (4):  Absolute $B$-magnitude.  
 
\item[ ]
Column (5):  Flux density (mJy) at 8.4 GHz and $1\sigma $ error, 
or $3 \sigma $ upper limit.  
 
\item[ ]
Column (6):  Logarithm of the ratio of 8.4 GHz luminosity 
(W ${\rm Hz^{-1}}$) to optical luminosity (W ${\rm Hz^{-1}}$) 
averaged over the $B$ passband; $R_{8.4} = L_{8.4}/L_{B}$.  
 
\item[ ]
Column (7):  Logarithm of 8.4 GHz luminosity (W ${\rm Hz^{-1}}$),
$\log L_{8.4}$.  

\end{itemize}

\subsection{Optical}
\label{ssec:dataoptical}

Many existing spectra of LBQS quasars suffer from systematic light
loss in the blue due to several instrumental effects and from poor
signal-to-noise ratio.  Higher quality long slit CCD spectra were
obtained with the Multiple Mirror Telescope (MMT) for 10 radio-loud
LBQS quasars on the night of UT 1995 Feb 5.  The new data, which cover
a wavelength range $3400 - 6900$ \AA, were reduced using standard
procedures.  The normalization of the flux scale is unreliable, due to
a narrow slit and thin cirrus during the observations, but the
relative flux calibration is unaffected, since the long axis of the
slit was maintained parallel to the direction of atmospheric
dispersion.

Improved spectra were not obtained for three radio-loud quasars with
$M_B$ near $-24$ (1430$-$0046, 2348$+$0210, and 2351$-$0036), so 
archival spectra were analyzed.  The spectrum for the latter object
does not appear to suffer from the instrumental blue light loss mentioned
previously.  The other two archival spectra show downturns blueward of
4000 \AA.  This wavelength region was excluded from spectral index
determinations. 

A spectral slope was determined for each of the 13 quasars using the
ratios of fluxes measured in pairs of fixed rest frame continuum
windows chosen to avoid prominent emission lines
(Francis\markcite{Francis93} 1993).  Only one pair of windows was
observable in each spectrum, either 2150$-$2230 to 3020$-$3100 \AA $ $
or 3020$-$3100 to 4150$-$4250 \AA $ $ rest frame, depending on the
redshift of the quasar.  Uncertainties in the spectral slopes were
calculated from the error in the mean flux of each window (standard
deviation across the window divided by the square root of the number
of spectral resolution elements) and a wavelength uncertainty equal to
half of the width of the window.  Equivalent widths were measured for
MgII and, for higher redshift objects, CIII], the only prominent
emission lines in the spectra of these quasars.

Broadband optical magnitudes were obtained with the Steward
Observatory 2.3 meter (nights of UT 1995 Feb 7 \& 8, Apr 8) and 1.5
meter (nights of UT 1995 Feb 19, Mar 14) telescopes.  Exposure times
were 5 minutes or less per frame.  In almost all cases, each quasar
was observed more than once per passband.  Standard reduction
techniques of bias subtraction and flat fielding were applied to all
frames, followed by the determination of instrumental magnitudes for
standard stars and program quasars using Stetson's
(1987\markcite{Stetson87}, 1990\markcite{Stetson90}) DAOPHOT and
DAOGROW packages.  Additional data, flagged in Table
\ref{tab:tableopt}, were drawn from a large multiband photometric
study of the LBQS, currently in preparation.

The $R$ passband is roughly coincident with restframe $B$ at redshifts
$z \sim 0.5$, making it more accurate to calculate $M_B$ from the $R$
magnitude, rather than observed $B$.  This conversion can be expressed
as
\begin{equation}
M_B - R = -2.5\left[\log \left(\frac{(d_{L}/10 {\rm pc})^{2}}{(1 +
z)^{\alpha + 1}}\right) + \log (\bar{F}_{B}/\bar{F}_{R}) + \log
(f_{\nu}(R = 0)/f_{\nu}(B = 0))\right],
\label{eq:RtoMB}
\end{equation} 
where $\bar{F}_{B}$ and $\bar{F}_{R}$ are the flux densities of a
power law spectrum with index $\alpha$ averaged over the restframe $B$
and $R$ passbands, respectively, $d_{L}$ is the luminosity distance,
and $f_{\nu}(B \mbox{ or } R = 0)$ is the flux density for magnitude
zero.  The error in the conversion from $R$ to $M_B \!\!$, given an
uncertainty in spectal index $\sigma_{\alpha}$, is
\begin{equation}
\sigma_{(M_{B} - R)} = \frac{2.5 \sigma_{\alpha}}{\ln 10}\left[\ln (1 +
z) - \frac{\int \nu^{\alpha} \ln \nu B(\nu)d\nu}{\int \nu^{\alpha}
B(\nu) d\nu} + \frac{\int \nu^{\alpha} \ln \nu R(\nu)
d\nu}{\int \nu^{\alpha} R(\nu) d\nu} \right].  
\label{eq:RtoMBerr}
\end{equation}
$B(\nu)$ and $R(\nu)$ are the filter transmission functions.  The
factor $\ln \nu$ arises from the derivative of $\nu^{\alpha}$ with
respect to $\alpha$.  Spectral indices determined for each quasar were
used in these calculations.  Absolute magnitude was derived from
observed $V$ in an analogous manner for the two quasars in Table
\ref{tab:tableopt} without $R$ magnitudes.  Uncertainties in the
conversion from $V$ or $R$ to $M_B$ were typically 0.03 mag and did not
exceed 0.05 mag for any object, leaving variability as the largest
error component for the 13 quasars with new photometry.  The final
errors in $M_B \!\!$, listed for each object in Table
\ref{tab:tablerad}, are a factor of 2 lower for the quasars with
improved optical data, compared to other LBQS quasars at the same
redshifts.  Absolute magnitudes for LBQS quasars other than those
listed in Table \ref{tab:tableopt} were calculated using the method
described in Paper II\markcite{PaperII}.

The newly obtained optical data, including one sigma errors, for
selected radio-loud quasars are presented in Table \ref{tab:tableopt}
as follows:

\begin{itemize}

\item[ ]
Column (1):  Object name.

\item[ ]
Columns (2) through (4):  $V$ magnitude, $B - V$ and $V - R$ 
colors, where
available.    

\item[ ]
Column (5):  Spectral index, $\alpha$ ($f_{\nu} \propto \nu^{\alpha}$).

\item[ ]
Columns (6) and (7):  Observed-frame equivalent widths for Mg II
$\lambda 2798$ and CIII] $\lambda 1909$.

\end{itemize}

\pagebreak

\section{Distribution of Radio Luminosity and \boldmath $ \log
R_{8.4}$ as Functions of $M_{B}$ \unboldmath} 
\label{sec:MB}

Figure \ref{fig:LnR_MB} shows the distributions of $\log L_{8.4} $ and
$\log R_{8.4} $ vs. $M_B$ for the data from Table \ref{tab:tablerad} plus
those from Table 2 in Paper II.  Detections with $\log L_{8.4} \geq 25$
or alternatively $\log R_{8.4} \geq 1$ are considered to be radio loud,
the same definitions employed in Paper II\markcite{PaperII}.  Quasars
in the new radio sample form a cluster of upper limits and detections
with elevated radio luminosities in the magnitude range $-25 < M_B <
-23$, due to shorter integration times than the observations described
in Paper II.

Radio-loud LBQS quasars are distributed fairly uniformly throughout a
range of over two decades in $R_{8.4}$ and $L_{8.4}$ and across all absolute
magnitudes sampled by the survey.  No correlation between either $\log
R_{8.4} $ or $\log L_{8.4} $ and $M_B$ is observed among the radio-loud
quasars, consistent with previous studies by Miller et
al.\markcite{MPM} (1990) and Stocke et al.\markcite{Stocke92} (1992).
The apparent correlation between radio luminosity and absolute
magnitude at low values of $\log L_{8.4} $ is the result of the LBQS flux
limit.

Comparisons between the LBQS and other optically selected surveys,
such as the PG, demonstrate the effects of different optical selection
criteria on the inferred radio properties.  Contrasting the LBQS with
surveys selected at other wavebands, such as the X-ray selected EMSS,
provides further insights into the relevant physical mechanisms and
helps identify radio properties of quasars which are universal,
regardless of selection frequency.  The overall radio-loud ($\log L_{8.4}
> 25$) fraction for the LBQS, 32/359 (9\%), is nearly identical to
that in the EMSS, 39/411 (9\%), but is less than half of the value for
the PG, 24/114 (21\%).  Overall fractions with $\log R_{8.4} > 1$ are
similar, with the LBQS at 31/359 (9\%), the EMSS at 46/411 (11\%), and
the PG at 22/114 (19\%).  Radio data for all of the PG quasars (Table
1 in Kellermann et al.\markcite{Kellermann89} 1989) and the EMSS
detections classified as ``AGN'' and having radio fluxes (Table 4 in
Stocke et al.\markcite{Stocke91} 1991) are used in the comparisons
throughout this paper.  Radio fluxes and luminosities at other
frequencies taken from the literature were extrapolated to 8.4 GHz
using a spectral index of $-0.5$, the same value employed for the LBQS
data.  The radio luminosity and $\log R_{8.4} $ distributions vs. absolute
magnitude for the LBQS differ from the PG and the EMSS, as shown in
Figure \ref{fig:smoothMB}.  These plots were constructed by
calculating the radio-loud fraction of objects within a 1 magnitude
range centered on the $M_B$ value of each quasar in the sample,
producing essentially a boxcar smoothing of the data.  The error bars
($\pm 1 \sigma$) shown at selected values of $M_B$ are the standard
deviations of a fraction calculated from a binomial random variable:
$\sigma = [f(1 - f)/N]^{1/2}$, where $f$ is the radio-loud fraction in
a 1.0 magnitude bin centered on $M_B$ containing $N$ points.  The
radio-loud fraction in the LBQS is constant at $\approx 10\%$ for all
but the most optically luminous quasars, rising to $\sim 35\%$ ($\sim
20\%$ for $\log R_{8.4} = 1$) for $M_B \gtrsim -28$.  The latter
fraction is rather uncertain due to the small number of objects in
this range.

A series of statistical tests were performed on the $M_{B}$
distributions of the radio-loud and radio-quiet quasars in the LBQS to
quantify the significance of the observed trends.  A Wilcoxon
two-sample rank sum test indicated that the radio-loud and radio-quiet
quasars do not have significantly different absolute magnitude
distributions (the statistic differed by only $0.03\sigma$ from the
value expected under the null hypothesis of equivalent $M_{B}$
distributions) using the $\log R_{8.4} > 1$ definition of radio-loud.
There is some indication of a difference using the $\log L_{8.4} > 25$
criterion (97\% confidence).  However, when the 11 quasars with $M_{B}
< -28$, the approximate region where the fraction of detections with
$\log L_{8.4} > 25$ is observed to rise, were excluded, no significant
difference (82\% confidence level) was found between the radio-loud
and radio-quiet distributions.  The Wilcoxon test was also applied to
the absolute magnitude distributions in a series of redshift bins in
order to remove some of the degeneracy between $z$ and $M_{B}$
inherent in any flux-limited sample.  Data for the 40 high-redshift
quasars in Schmidt et al.\markcite{Schmidt95} (1995), discussed in
more detail in Section \ref{sec:z}, were added to the LBQS radio
sample to extend the redshift range.  Two binning schemes were
employed, one using the redshift ranges of Schmidt et
al.\markcite{Schmidt95} (1995), shown in Figure \ref{fig:LBQScum}, and
in the other case the combined sample was divided into 5 bins of
nearly equal numbers.  The only moderately significant (96.8\%
confidence) difference between the radio-loud, using either
definition, and remaining distributions occured for $\log L_{8.4} >
25$ in the equal-number bin covering the redshift range $2.34 < z <
4.90$.  Again, this was due to the enhanced radio-loud fraction at
bright absolute magnitudes.  Ties in the rankings for the Wilcoxon
tests were handled in the standard way by averaging the ranks of a
group of quasars with the same $M_{B}$, and the probabilites quoted
are for a two-sided test using the asymptotic normal distribution of
the rank sum statistic.

The LBQS radio data, sorted by $M_{B}$, were divided into 10 bins each
containing $\approx 36$ objects to perform some of the $\chi^{2}$ and
binomial tests described in \S3.2.2 of Paper II\markcite{PaperII}.
The results are very close to those in Paper II\markcite{PaperII} and
are consistent with the results of the Wilcoxon tests above.  The
number of detections with $\log R_{8.4} > 1$ in each bin is consistent
with a constant radio-loud fraction equal to the average fraction for
the whole LBQS radio sample (the probability that the sample is not
consistent with the flat distribution is 81\%).  This is not the case,
however, using the radio luminosity criterion for radio-loud, which
produces a distribution which is not flat at the 99.7\% confidence
level.  The discrepency lies in the brightest absolute magnitude bin,
$-28.7 < M_{B} < -27.5$, and if this bin is excluded, the confidence
level drops to an insignificant 76\%.  The liklihood from a binomial
distribution of selecting at least the observed number, 10/36 (28\%),
of radio-loud quasars in the brightest bin, given the average
radio-loud fraction for the full sample, is 0.00092.  Since this bin
was selected a posteriori because of its high incidence of radio-loud
quasars, the true significance of the enhanced radio-loud fraction is
the binomial probability multiplied by the number of bins (10), for a
final confidence level of 99.1\%.

All of the statistical tests performed on the distribution of
radio-loud objects with $M_{B}$ in the LBQS sample are consistent with
the representation shown in Figure \ref{fig:smoothMB}a: the radio-loud
fraction is essentially constant except at the brightest absolute
magnitudes, and even then it is only significantly higher using the
$\log L_{8.4} > 25$ criterion.  Some possible interpretations of this
distribution are discussed in Section \ref{sec:discussion}.

Both the PG and EMSS have substantially higher radio-loud fractions
than the LBQS for $M_B$ brighter than $-24$ (see Figure
\ref{fig:smoothMB}, b \& c).  Note that the PG has an
anomolously high radio-loud fraction in this absolute magnitude range
compared to other optically selected quasar samples (La Franca et
al.\markcite{LaF94} 1994; Della Ceca et al.\markcite{Della94} 1994).
The reason for this discrepency remains unexplained.  A strong trend
in radio-loud fraction as a function of both absolute magnitude and
redshift is expected in the EMSS, due to the well-known correlations
between radio, X-ray, and optical luminosity (see, e.g., Worrall et
al.\markcite{Worrall87} 1987) and the X-ray flux limit.  Figure
\ref{fig:zMB} shows absolute magnitude plotted against redshift for
the LBQS and EMSS, with the radio-loud ($\log L_{8.4} > 25$) quasars
indicated by filled triangles.  It is apparent that the observed
trends with redshift and absolute magnitude are mostly degenerate.  In
particular, all but one of the radio-loud EMSS quasars brighter than
$M_B = -24$ have $z > 0.5$, whereas the bulk of the radio-quiet
quasars fainter than this absolute magnitude are at redshifts $z <
0.5$.  However, some discrimination as a function of $M_B$ is possible
in the redshift interval $0.4 < z < 0.7$.  In this redshift range the
radio-loud fraction is 7/20 (35\%) for $M_B < -24$ and 3/45 (7\%)
among the fainter quasars, suggesting that at $z \sim 0.5$ the
radio-loud fraction of X-ray selected quasars is dependent upon
absolute magnitude, decreasing for $M_B$ fainter than $-24$.  The
selection effect proposed by Peacock et al.\markcite{Peacock86} (1986)
does not explain the decline in radio-loud fraction for $M_B > -24$
in the EMSS, since the ratio of quasar to host galaxy optical
luminosity is not expected to affect X-ray selection.

The Peacock et al.\markcite{Peacock86} (1986) selection effect is
still a viable explanation for the decrease in radio-loud fraction in
the PG, since a quantitative test of the applicability of this
hypothesis to the PG cannot be performed, as was done for the LBQS in
Paper II.  Alternatively, the PG may have a nearly normal radio-loud
fraction for $M_B > -24$, perhaps only slightly depressed by the
selection effect, which only appears abnormally low because the
fraction at brighter absolute magnitudes is atypically high.  If the
radio-loud fraction among the optically faint PG quasars is compared
to the fraction at absolute magnitudes brighter than $-24$ in the
LBQS, rather than in the PG, the significance of the change becomes
equivocal.  The fraction of quasars with $\log R_{8.4} > 1$ in the faint
PG sample (2/45, 4\%) is not significantly different (71\% confidence
level) from the LBQS value of 23/286 (8\%) for $M_B < -24$.  However,
the difference is significant at the 98.6\% confidence level using the
$\log L_{8.4} $ definition of radio-loud, for which the fractions are
0/45 and 26/286 (9\%) in the PG and LBQS subsamples, respectively.  If
PG quasars with $M_B$ fainter than $-22.7$, the faint absolute
magnitude limit of the LBQS, are excluded from the comparison, the
fractions for both definitions become 0/24, for which the confidence
levels are 87\% ($\log R_{8.4} > 1$) and 90\% ($\log L_{8.4} > 25$).

No pronounced decline in radio-loud fraction at absolute magnitudes
fainter than $-24$ is seen in the LBQS, contrary to the other two
samples (Figure \ref{fig:smoothMB}).  As noted in the introduction, a
decrease was seen in this range in Paper II\markcite{PaperII}, but the
number of objects with $M_B > -24$ was small (20), and the absolute
magnitude uncertainties were $\pm 0.4$ mag.  The new LBQS data
increase the sample size at faint optical luminosities by more than a
factor of 3 and decrease the magnitude uncertainties for radio-loud
objects near $M_B = -24$ by a factor of 2.  Improved absolute
magnitude determinations for two radio-loud quasars listed as slightly
brighter than $M_B = -24 $ in Paper II\markcite{PaperII} (see Figure
4 of that work) now place them fainter than this division, along with
4 new radio-loud quasars.  The radio-loud ($\log L_{8.4} > 25$) fraction
in the combined LBQS sample is 6/73 (8\%) at absolute magnitudes
fainter than $M_B = -24$, nearly identical to the fraction (26/286,
9\%) at brighter absolute magnitudes.  Much of the discussion to date
on the radio luminosity distribution of quasars at low optical
luminosities has been based at least in part on the PG (Peacock et
al.\markcite{Peacock86} 1986; Miller et al.\markcite{MPM} 1990;
Padovani\markcite{Padovani93} 1993; Falcke, Gopal-Krishna, \&
Biermann\markcite{Falcke95} 1995).  The LBQS results show that the
radio-loud fraction does not decline for $M_B > -24$ in all optically
selected quasar samples.  In cases where it does, the cause may be a
selection effect at faint absolute magnitudes or an increased
prevalence of radio-loud quasars at bright absolute magnitudes.

\pagebreak

\section{The Evolution of Radio and Optical Luminosity}
\label{sec:z}

Radio-loud fraction is nearly constant in the LBQS over most of the
redshift range of the survey, as shown in Figure
\ref{fig:smoothzLBQS}.  The smoothed fraction as a function of $z$ was
calculated with a redshift smoothing interval of $0.5$ in a manner
analogous to its determination as a function of $M_B \!\!$, above.
Some of the same statistical tests used in Section \ref{sec:MB} were
applied to the distributions of $\log L_{8.4}$ and $\log R_{8.4}$,
with a similar result that the quantitative measures confirm the
trends displayed graphically in Figure \ref{fig:smoothzLBQS}.  The
Wilcoxon rank sum statistic for detections with $\log R_{8.4} > 1$
differs by only $0.2\sigma$ from the value expected if these have the
same redshift distribution as the remaining quasars in the LBQS radio
sample.  The radio-loud quasars have a moderately significantly
(confidence level 95\%) different redshift distribution using $\log
L_{8.4} > 25$, but the significance disappears (revised confidence
level 77\%) when the region with the observed spike in radio-loud
fraction, $z > 3$, is excluded.  The results of a $\chi^{2}$ goodness
of fit test of a model in which the radio-loud fraction is unevolving
and equal to the sample average are not as clear-cut as those from the
Wilcoxon test, because of a greater sensitivity to the rise in
radio-loud fraction around $z \approx 1$.  While the increase in
radio-loud fraction for $z \approx 1$ is not nearly as great as the
apparent climb at $z \geq 3$, this enhancement contributes
significantly to the $\chi^{2}$ statistic.  For example, the $\log
R_{8.4}$ distribution, which has very close to the expected number of
radio-loud quasars in the highest redshift bin, is inconsistent with a
flat distribution at the 94\% confidence level based on the $\chi^{2}$
test.

A model for the evolution of the radio-loud fraction was constructed
from the radio luminosity functions of Dunlop \&
Peacock\markcite{Dunlop90} (1990).  They derived separate luminosity
functions for flat and steep spectrum objects, divided at a spectral
index $\alpha = -0.5$, from a radio-selected sample of quasars and
radio galaxies.  Distinct versions of the model were formed from the
steep and flat spectrum RLF1 MEAN-$z$ luminosity functions, since
radio spectral information is not available for most quasars in the
LBQS.  The luminosity functions were truncated at a radio luminosity
equivalent to $\log L_{8.4} = 28.85$, two decades below the limit
employed by Dunlop \& Peacock\markcite{Dunlop90} (1990).  The steep
spectrum RLF1 function has an unphysical sharp upturn in number
density at luminosities greater than this value.  Function RLF2
remedies this with an imposed exponential cut-off at high
luminosities, but the model based on this luminosity function does not
match the observed data as well as the truncated RLF1.  The true
number of objects excluded by truncating the luminosity functions
nearly 4 orders of magnitude above the radio-loud threshold of $\log
L_{8.4} = 25$ is expected to be relatively small and have negligible
impact on the predicted space density of radio-loud quasars.  The
$\log R_{8.4}$ radio-loud criterion was also employed, although
indirectly, as the luminosity functions do not provide a one-to-one
correspondence between radio luminosity and absolute magnitude.  A
radio luminosity threshold which depends on redshift, $\log L_{8.4} =
24.6 + 0.36z$, was used to approximate the $\log R_{8.4} = 1$ boundary
(see Figure 7 in Paper II).  The normalization of the predicted
fraction is set by the overall radio-loud fraction of the LBQS, as
described, along with other details of the model, in Paper II.  The
predicted evolution of radio-loud fraction for various model
parameters is shown with the observed smoothed fraction in Figure
\ref{fig:smoothzLBQS}.  The flat spectrum versions match the data
well, including the modest rise around $z = 1$, at all except the
highest redshifts, where the model does not reproduce the substantial
rise in the fraction with $\log L_{8.4} > 25$.  A similar local maximum
occurs in the steep spectrum version, but at higher redshifts than
observed.

Many of the quasars at high redshift, where the radio-loud fraction
rises, have absolute magnitudes $\sim -28$ (Figure
\ref{fig:smoothMB}a).  It is not possible to disentangle redshift from
$M_B$ as the causal variable for the enhanced radio-loud fraction using
only the LBQS data.  However, the radio-loud fraction in the
restricted redshift ($1.8 < z < 2.5$) sample of Miller et
al.\markcite{MPM} (1990) rises steeply for $M_B$ brighter than $-27$,
indicating that elevated radio-loud fraction at bright absolute
magnitudes is a common feature of optically selected quasars over a
wide range in redshift.  A similar ambiguity between $M_B$ and $z$
involving the rise at $z = 1$ was discussed in Paper
II\markcite{PaperII}, where it was concluded that the effect is
probably evolutionary.  The new data support this, as there is no peak
in radio-loud fraction at $M_B \sim -25.5$, the typical absolute
magnitude for LBQS quasars with $z \sim 1$ (Figure
\ref{fig:smoothMB}).

Optically selected quasars with $z > 3.0$ have a radio-loud fraction
similar to that of the LBQS.  McMahon, Irwin, \&
Hazard\markcite{McMahon92} (1992) collected radio data for 29 quasars
in the redshift range $3.5 < z < 4.7$, 4 of which have $\log L_{8.4} >
25$, and 3 have $\log R_{8.4} > 1$.  Individual absolute magnitudes and
redshifts for these quasars were not published.  Schneider et
al.\markcite{Schneider92} (1992) observed 22 quasars at 5 GHz, which
were reobserved at 1.5 GHz along with an additional 18 quasars by
Schmidt et al.\markcite{Schmidt95} (1995).  One target was detected in
the original study, and it was also above the detection threshold at
1.5 GHz, along with two others which were not included in the first
sample.  The common detection has $\log L_{8.4} > 25$ and $\log R_{8.4} > 1$
for both observations.  The remaining two low frequency detections
have $\log L_{8.4} > 25$, but only one is radio-loud by the $R_{8.4}$
criterion.  The 40 quasars in the combined sample lie in a redshift
interval $3.1 < z < 4.9$ and have absolute magnitudes ranging from
$-24.1$ to $-27.8$, with an average of $-26.3$ (values taken directly
from Schmidt et al.\markcite{Schmidt95} 1995).  Merging the high-$z$
data sets gives a high redshift radio-loud fraction of 7/69 (10\%) for
$\log L_{8.4} > 25$ and 5/69 (7\%) using the luminosity ratio
criterion, values consistent with the LBQS in the redshift range $1.5
< z < 2.5$.  A Wilcoxon test on the combined LBQS and Schmidt et
al.\markcite{Schmidt95} (1995) samples found no statistically
significant (88\% confidence level) difference in the redshift
distributions of the detections with $\log L_{8.4} > 25$ and the
remaining quasars, even without removing the high-$z$ LBQS objects.
Similar results were obtained when this combined sample was divided
into 5 equal-number absolute magnitude bins.  No significant
differences in the redshift distributions were found among the bright
bins, which contain the high-$z$ quasars.  Ten of the quasars in the
Schmidt et al.\markcite{Schmidt95} (1995) sample lie in a redshift
range which overlaps the LBQS ($3.1 < z < 3.4$), and none are radio
loud.  All of these have $M_B$ fainter than $-27.3$, lending further
support to the conclusion that the apparent rise in radio-loud
fraction at high $z$ in the LBQS is a manifestation of a trend with
absolute magnitude.  Therefore, despite the appearance of Figure
\ref{fig:smoothzLBQS} at high redshift, there is substantial 
evidence that the radio-loud fraction of optically selected quasars
does not evolve, aside from a modest peak around $z \approx 1$, over
65\% of cosmic time, from $z = 0.2$ to redshifts approaching 5: (1)
the radio-loud fraction at high $z$ is consistent with that of the
LBQS at moderate redshift; (2) a model which matches the LBQS data
well at most redshifts does not reproduce the large fraction at high
$z$; and (3) $M_B$ is the causal variable for the apparent rise in
radio-loud fraction for $z \gtrsim 3$ in the LBQS, indicated by higher
redshift samples.  The overall radio-loud fraction in the high-$z$
sample and the LBQS data with $z < 3.1$, together covering the
redshift range $0.2 < z < 4.9$, is 37/426 (9\%) for $\log L_{8.4} > 25$
and 34/426 (8\%) for $\log R_{8.4} > 1$.

A nearly unevolving radio-loud fraction is inconsistent with many
previous studies, which compared the predominantly low-$z$ PG with
higher redshift optically selected samples and found a pronounced
decrease in radio-loud fraction with increasing redshift (Peacock et
al.\markcite{Peacock86} 1986; Miller et al.\markcite{MPM} 1990; Paper
I\markcite{Visnovsky92}; Schneider et al.\markcite{Schneider92} 1992).
La Franca et al. (1994) noted the same trend but found that evolution
of the fraction was no longer indicated conclusively if the PG were
excluded from the analysis.  Recently, Schmidt et
al.\markcite{Schmidt95} (1995) compared cumulative distributions of
$\log R$ in 4 non-overlapping redshift ranges using data from a
different survey for each distribution.  The cumulative functions for
the 3 highest redshift bins ($z > 1$) do not indicate substantial
evolution over this range, but the low redshift distribution ($z <
0.6$), containing only PG quasars, has a significantly larger fraction
of quasars with high values of $\log R$, especially for $M_B < -24.5$
(see their Figures 2 \& 3).  They concluded that the radio properties
of quasars have evolved such that either the values of $R$ have
decreased by two orders of magnitude from $z = 0.3$ to 1.3 while
maintaining the same distribution or that a population of quasars with
very weak radio emission was more prevalent at $z > 1$.

The redshift coverage of the LBQS is extensive enough to generate
cumulative distributions in 3 of the 4 redshift bins from Schmidt et
al.\markcite{Schmidt95} (1995) from a single well-defined survey, with
the following constraints.  The LBQS has little overlap with their
highest redshift bin ($3.1 < z < 4.9$), in which they used their own
observations.  Many of the PG quasars in the low-$z$ bin ($z < 0.6$)
have $z < 0.2$, the lower redshift limit of the LBQS.  The absolute
magnitude discrimination used for most of the analysis by Schmidt et
al.\markcite{Schmidt95} (1995), $M_B < -24.5$, excludes almost 90\%
of these low-$z$ PG quasars but also restricts the LBQS to $z > 0.35$.
Redshift-segregated cumulative distributions of LBQS quasars are shown
in Figure \ref{fig:LBQScum}, along with the high-$z$ data of McMahon
et al.\markcite{McMahon92} (1992) and Schmidt et
al.\markcite{Schmidt95} (1995) plotted as a single distribution.
Kolmogorov-Smirnov tests were performed on these distributions in two
ways.  In one case, the true $\log R_{8.4}$ values of the upper limits
were assumed to all lie below the smallest value for a detection, as
displayed in Figure \ref{fig:LBQScum}.  Only the portions of the
distributions consisting of detections were compared to derive the
test statistic.  In the other case, the cumulative distributions were
formed only from the detections, ignoring the upper limits completely.
A significant difference was not found in any of these tests, the
highest confidence level being 92\%.  The cumulative distributions in
the four redshift ranges are similar, indicating no evolution.  In
particular, the $z < 0.6$ distribution, which had a cumulative
fraction substantially higher than the other redshift ranges in the
Schmidt et al.\markcite{Schmidt95} (1995) analysis, is slightly below
that for $1.0 < z < 1.8$ for most values of $\log R_{8.4} $.

While the radio-loud fraction in the LBQS shows little dependence on
redshift, the EMSS exhibits a rapid rise with redshift until reaching
a plateau at $\approx 40\%$ for $z > 1$, as shown by the solid line in
Figure \ref{fig:smoothzEMSS}.  The fractions with $\log L_{8.4} > 25$
and $\log R_{8.4} > 1$ are nearly identical; the latter is slightly
higher at low redshifts.  Note that the behavior of the radio-loud
fraction at the high redshift end of this plot is particularly
uncertain, as there are only 4 quasars in the EMSS sample with $z >
2$.  This trend of radio-loud fraction with redshift is expected in
X-ray selected samples as a result of the correlation between X-ray
and radio luminosities, which is thought to arise from a synchrotron
self Compton component of the X-ray emission.  At higher redshifts the
flux limits of an X-ray survey correspond to higher luminosities,
which, because of the correlation, tend to have higher radio
luminosities.

A model for the relationship between radio-loud fraction and redshift
in the EMSS was constructed using the expression for the dependence of
X-ray luminosity on radio luminosity given by Brinkmann et
al.\markcite{Brinkmann95} (1995).  They parameterized the correlation
as $\log (l_{x}) = A_{\rm rq} + \beta_{\rm rq} \times \log (l_{r})$,
where $l_{x}$ and $l_{r}$ are X-ray and radio monochromatic
luminosities at 2 keV and 5 GHz, respectively.  The constants were
determined from ROSAT All-Sky Survey (Voges\markcite{Voges92} 1992)
data on quasars in the 87 Green Bank radio survey (Condon et
al.\markcite{Condon89} 1989) to be $A_{\rm rq} = 8.44 \pm 0.40$, and
$\beta_{\rm rq} = 0.56 \pm 0.03$, where the uncertainties quoted here
have been converted to $1\sigma$ errors assuming Gaussian probability.
Rest frame 2 keV luminosities were determined from the survey flux
limit at each redshift using a spectral index $\alpha = -1.13$
(Brinkmann et al.\markcite{Brinkman95} 1995).  The corresponding value
of $l_{r}$ was used as the lower luminosity limit in the integrals of
the Dunlop \& Peacock\markcite{Dunlop90} (1990) steep and flat
spectrum MEAN-$z$ RLF1 luminosity functions.  The upper limit was the
same as for the LBQS comparison model.  These functions were also
integrated over a redshift interval of 0.5 to find the total expected
number of quasars.  The number of radio-loud quasars was determined by
integrating the same function using as the lower limit the larger of
the radio-loud luminosity threshold, converted to the units employed
in the luminosity function, or the luminosity determined from the
correlation.  This threshold was either $\log L_{8.4} = 25$ or the
redshift-dependent value equivalent to $\log R_{8.4} = 1$ discussed
earlier.  The radio-loud boundary was not permitted to be lower than
the luminosity corresponding to the 0.9 mJy flux limit of the radio
observations of the EMSS (Stocke et al.\markcite{Stocke91} 1991).  The
EMSS did not have a uniform X-ray flux limit.  A weighted average for
each integral was calculated from 6 values of limiting X-ray flux
determined from Gioia et al.\markcite{Gioia90} (1990), with the
weights proportional to the number of EMSS objects found at each
limiting flux.

The predicted overall EMSS radio-loud fractions range mostly between
10\% and 20\% (cf. the observed fractions of 9\% ($\log L_{8.4}$) and
11\% ($\log R_{8.4}$) -- Section \ref{sec:MB}) as the values of
$A_{\rm rq}$ and $\beta_{\rm rq}$ were varied within their $1\sigma$
uncertainties.  Values for these parameters were chosen so that the
predicted average radio-loud fraction for the EMSS sample matched that
observed.  Each combination of radio-loud definition and luminosity
function in the model was normalized separately.  The predicted
radio-loud fractions are plotted in Figure \ref{fig:smoothzEMSS},
along with the observed fractions.  Most of the EMSS quasars are at
low redshift (median $z$ is 0.268), where the radio-loud fraction is
small, resulting in the modest overall radio-loud fraction.  While
there is little difference between the steep and flat spectrum models,
the form of the predicted redshift dependence varies widely within the
$1\sigma$ uncertainties of the correlation parameters, and none of the
model distributions match the observed one for $z > 1$.  The excess of
predicted radio-loud quasars at high redshifts and the relative
depletion at lower $z$ are directly connected, due to the
normalization of the model by the observed average fraction.  Radio
luminosity has a spread of almost 2 dex for all $l_{x}$ in the
Brinkmann et al.\markcite{Brinkmann95} (1995) sample, which would
lower the predicted radio-loud fraction, especially at high redshift.
In particular, the model fraction would not reach 100\% until a
redshift at which the radio luminosity predicted from the X-ray
limiting flux attained a value $\sim 1$ dex above the radio-loud
boundary, since the intrinsic scatter in $l_{r}$ would provide some
radio-quiet quasars up to this point.  While these simple models do
not conform to the observed data over all redshifts, they indicate
that the rise in radio-loud fraction for $z < 1$ is an expected
feature of the X-ray/radio luminosity correlation.

\pagebreak

\section{Discussion}
\label{sec:discussion}

Understanding the connection between the radio emission and the
properties of the central engines of quasars continues to be an
elusive problem.  The constancy of the radio luminosity distribution
over more than two orders of magnitude in optical luminosity and other
similarities between radio-loud and radio-quiet quasars are somewhat
suprising in the context of the standard black hole model of AGN.  The
radio emission mechanism appears to be only weakly tied to the mass
accretion rate which presumably drives the optical and near-UV
emission.  Some models decouple the radio and optical emission, an
important feature in attempting to explain the results from the LBQS
and other studies which show little connection between these emission
mechanisms.  Coleman \& Dopita \markcite{Coleman92} (1992)
hypothesized that quasar radio luminosity may depend on the angle of
the accretion disk with respect to the rotation axis of the central
supermassive black hole.  More recently however, Wilson \&
Colbert\markcite{Wilson95} (1995) proposed that the hole masses and
accretion rates are similar in radio-loud and quiet quasars, which
would account for the similarities in emission features at other
wavebands.  The total radio luminosity is presumed to derive from the
spin energy of the hole via the Blandford-Znajek process (Blandford \&
Znajek\markcite{Blandford77} 1977).  Powerful radio emission arises
only from massive rapidly rotating holes, which in the view of Wilson
\& Colbert are produced exclusively by mergers of two holes with
similar large masses.  They assume that accretion of material from the
host galaxy is insufficient to spin up the hole to the necessary level
because of either loss of angular momentum to gas around the hole or
variable orientation of the accretion disk with respect to the spin
axis of the hole.  Therefore, two quasars with identical black hole
masses and accretion rates can have vastly different radio
luminosities depending on whether the hole was 1) built up by
accretion and/or the merger of a large and small hole or 2) produced
by the coalescence of two massive holes.  The infrequency of mergers
of galaxies containing massive black holes accounts for the rarity of
powerful radio sources.  The Wilson \& Colbert scenario is
qualitatively consistent with the LBQS results that radio luminosity
is not correlated with $M_B$ and that the radio-loud fraction is
constant over most values of $M_B \!$.

A different radio emission mechanism which {\it is} correlated with
optical luminosity may be responsible for the rise in radio-loud
fraction at bright absolute magnitudes ($M_B \lesssim -28$), seen
also by Miller et al.\markcite{MPM} (1990) and La Franca et
al.\markcite{LaF94} (1994).  Stocke et al.\markcite{Stocke92} (1992)
noted that radio and optical luminosity are correlated among the
radio-quiet PG quasars.  The detection frequency among the radio-quiet
objects in the LBQS is too low to directly test for such a
correlation, but there is indirect evidence that such a correlation is
present: the fraction of radio detections among the radio-quiet LBQS
quasars does not show a trend with absolute magnitude.  If there were
a null or negative correlation between the luminosities, the detection
fraction among these objects would decline at higher redshifts,
corresponding to brighter $M_B \!$, since the radio luminosity of the
$3 \sigma $ detection limit increases with $z$.  Stocke et
al.\markcite{Stocke92} (1992) hypothesized that the correlation arises
from a component of radio emission powered by the kinetic energy of
outflowing material driven by optical/UV radiation pressure.
Radiation pressure and consequently outflow velocity aren't large
enough to produce $\log L_{8.4} > 25$ by this mechanism alone for $M_B
\gtrsim -28$, but this emission component could exceed the radio-loud
threshold ($\log L_{8.4} > 25$) at the brightest absolute magnitudes in
the LBQS.  In this scenario the enhanced radio-loud fraction at $M_B
\lesssim -28$ arises from the combination of the black hole
spin-driven component, which is independent of $M_B \!$, and the
onset of powerful radio emission driven by outflows.  The presence of
two radio emission mechanisms, one correlated with optical luminosity,
the other independent, accounts for the observed features of the radio
luminosity distribution as a function of $M_B$ in the LBQS.

Neither radio emission mechanism explains the low radio-loud fraction
observed at $M_B > -24$ in the EMSS and the PG surveys.  Falcke et
al.\markcite{Falcke95} (1995) proposed that the decrease in radio-loud
fraction at faint optical luminosities can be explained by a
consolidation of two standard quasar unification schemes incorporating
obscuring tori (e.g., Barthel\markcite{Barthel89} 1989; Urry et
al.\markcite{Urry91} 1991; Antonucci\markcite{SkiARAA} 1993).
Radio-loud AGN with weaker (presumably fainter $M_B \!$) engines are
observed as either BL Lac objects or FR I radio galaxies, not as
quasars, as the opening angle of the torus is not wide enough in these
objects for a direct view of the nuclear region free from
contamination by the jet emission.  This explanation is not directly
tenable for the LBQS, since the incidence of radio-loud quasars is not
observed to change at faint $M_B \!$.  It is possible that, due to
some poorly understood difference between the PG and the LBQS,
orientation effects are present in the LBQS at a weaker level than
originally proposed by Falcke et al.\markcite{Falcke95} (1995) to
explain the PG results.  If the lines of sight to the radio-loud
quasars with $M_B > -24$ lie close to relativistic jets (to avoid
obscuration by the tori), the emission lines are expected to have low
equivalent widths due to their dilution by doppler-boosted optical
synchrotron emission; i.e., these AGN would be BL Lac-like, but not so
much that they would be classified as BL Lacs.  However, the
radio-loud LBQS quasars with faint absolute magnitudes are not close
to being classified as BL Lacs, as 4/6 have prominent emission lines,
and the lines in the remaining two, while weaker, still have MgII
equivalent widths of approximately 15\AA $ $ (Table
\ref{tab:tableopt}).  AGN are classified as BL Lacs typically when
they have emission line equivalent widths $< 5$\AA $ $ (e.g., Stocke
et al.\markcite{Stocke91} 1991).  The model does not appear to be
applicable to optically selected quasar samples, as the proposed
orientation effect is not present in the LBQS at a detectable level,
and a reasonable alternative explanation exists for the PG, although
the model may partially account for the decrease observed in the EMSS.
 
\pagebreak

\begin{center}
\section{Summary}
\label{sec:summary}
\end{center}

\renewcommand{\labelenumi}{\roman{enumi}.}
\begin{enumerate}

\item  The distribution of radio luminosity does not depend on
absolute magnitude over most of the range of $M_B$ in the LBQS.  The
radio-loud fraction remains constant at $\approx 10\%$ for $-28
\lesssim M_B \lesssim -23$ but rises at brighter absolute 
magnitudes to $(20 \pm 10)\% $ for $\log R_{8.4} > 1$ and $(35 \pm
15)\% $ for $\log L_{8.4} > 25$.

\item  Radio-loud fraction is nearly unevolving at a value of $\approx
10\%$ in the LBQS and three high-$z$ optically selected samples, which
together span a redshift range from $z = 0.2$ to redshifts approaching
5.

\item 
The PG survey differs significantly from several optically selected
samples.  The sudden decrease in radio-loud fraction for $M_B$ fainter
than $-24$ observed in the PG is not present in the expanded LBQS
sample.  The fraction of quasars with $\log R_{8.4} > 1$ in the PG for
$M_B > -24$ is not significantly different from the LBQS fraction,
but the PG has a much greater percentage of radio-loud quasars at
brighter absolute magnitudes.
High radio-loud fraction in the PG relative to other optically
selected samples has led several investigators to conclude that
the fraction evolves to lower values at higher $z$.  This trend is not
seen in the LBQS, which provides a single well-defined sample with
wide redshift coverage ($0.2 < z < 3.4$).  The anomolously high
radio-loud fraction in the PG sample remains unexplained.    

\item  The radio properties of the X-ray selected EMSS also differ
from those of the LBQS.  The rapid rise in radio-loud fraction
observed for $M_B \approx -24$ arises primarily from a
well-established correlation between X-ray and radio luminosity.
However, there is still a dependence of radio-loud fraction on $M_B$ in
the redshift range $0.4 < z < 0.7$, an interval containing substantial
numbers of quasars both brighter and fainter than $M_B = -24$.

\item  The behavior of the radio properties of the LBQS quasars as a
function of $M_B$ is consistent with the existence of two radio
emission mechanisms, one correlated with optical luminosity, the other
independent.

\end{enumerate}

\acknowledgments
The authors thank Pat Hall for conducting some observations and Steve
Warren for helpful discussions.  The data would not have been obtained
without the capable and friendly help of the operations staffs at the
VLA, MMT, and Steward Observatory.  Commentary from an anonymous
referee led to clarification of the manuscript.  This work was
partially supported by NSF grant AST 93-20715 and NASA grant
NGT-51152, a NASA Graduate Student Researchers Program Fellowship
(EJH).  This research has made use of the NASA Astrophysics Data
System (ADS).

\pagebreak

\begin{center}
Figure Captions
\end{center}

\noindent
{\bf Fig. \ref{fig:zMBLBQS}} \hspace{0.5cm} Distributions of $M_{B}$
against redshift for the LBQS radio detections ({\it filled circles})
and upper limits ({\it horizontal lines}).  Y-shaped symbols represent
LBQS quasars without radio observations.  All of the latter class with
$z > 2.5$ are plotted, but the large numbers at lower redshifts would
over-clutter the plot.  Therefore, a random selection of 1/3 of the
quasars with $z < 2.5$ lacking radio data are plotted.
 
\noindent
{\bf Fig. \ref{fig:LnR_MB}} \hspace{0.5cm} The distribution of (a)
$\log R_{8.4} $ and (b) $\log L_{8.4} $ against $M_B \!$ for the LBQS.  The
data shown consist of the new observations from Table
\ref{tab:tablerad} and those from Table 2 in Paper
II\markcite{PaperII}.  Filled circles represent detections, and
horizontal lines are upper limits.  Error bars ($\pm 1\sigma$),
calculated as described in Section \ref{sec:data}, are shown for a
representative point at bright ($M_B = -27.8$; $\log R_{8.4} = 3.5$,
$\log L_{8.4} = 28.3$) and and one at faint ($M_B = -23.2$; $\log R_{8.4} =
1.3$, $\log L_{8.4} = 24.3$) absolute magnitudes and individually for
those quasars which were reobserved in the optical.  Error bars in
$\log L_{8.4} $ are not plotted in (b) for the quasars with new
photometry to avoid unnecessary crowding and because the radio
luminosity uncertainties are not affected by the new data.

\noindent
{\bf Fig. \ref{fig:smoothMB}} \hspace{0.5cm} Smoothed radio-loud
fraction as a function of $M_B$ (smoothed over an interval of 1.0 mag)
for the LBQS (a), the PG (b), and the EMSS (c) surveys.  Solid lines
represent the fraction with $\log L_{8.4} > 25$, and dotted lines are for
$\log R_{8.4} > 1$.  Error bars, calculated as described in the text, are
shown at selected values of $M_B \!$.

\noindent
{\bf Fig. \ref{fig:zMB}} \hspace{0.5cm} Absolute magnitude
vs. redshift for the LBQS (a) and the EMSS (b).  This is similar to
Figure \ref{fig:zMBLBQS}, but in this case the filled triangles are
radio-loud ($\log L_{8.4} > 25$) detections, crosses represent
radio-quiet detections, and horizontal lines indicate upper limits,
assumed to be radio-quiet.

\noindent
{\bf Fig. \ref{fig:smoothzLBQS}} \hspace{0.5cm} Smoothed radio-loud
fraction as a function of redshift in the LBQS sample using the (a)
$\log L_{8.4} > 25$ and (b) $\log R_{8.4} > 1$ criteria.  Solid lines
indicate the observed data, while dotted and dashed lines represent
the flat and steep spectrum model predictions, respectively.  Values
for the data as well as the models were calculated with a redshift
smoothing interval of 0.5.  Error bars, calculated in an analogous
manner to those in Fig. 3, are shown at representative data values.

\noindent
{\bf Fig. \ref{fig:smoothzEMSS}} \hspace{0.5cm} Smoothed radio-loud
fraction as a function of redshift in the EMSS sample using the (a)
$\log L_{8.4} > 25$ and (b) $\log R_{8.4} > 1$ criteria.  Solid lines
indicate the observed data, while dotted and dashed lines represent
the flat and steep spectrum model predictions, respectively.  Values
for the data as well as the models were calculated with a redshift
smoothing interval of 0.5.  The error bars are analogous to those in
the other plots of smoothed radio-loud fraction.

\noindent
{\bf Fig. \ref{fig:LBQScum}} \hspace{0.5cm} The cumulative
distributions $G ( > \log R_{8.4} )$ of LBQS quasars in three redshift
ranges: squares represent $z < 0.6$; filled circles $1.0 < z < 1.8$;
and plus signs $1.8 < z < 2.5$.  The high redshift range $3.1 < z <
4.9$, represented by stars, contains the samples of McMahon et
al.\markcite{McMahon92} (1992) and Schmidt et al.\markcite{Schmidt95}
(1995).  Panel (a) includes all quasars with $M_B < -23$, and the
absolute magnitude restriction is $M_B < -24.5$ in panel (b).  The
redshift bins, symbols, and absolute magnitude ranges match those used
in Schmidt et al.\markcite{Schmidt95} (1995).  Only the detections are
plotted for clarity.  All of the quasars in each redshift range were
used to calculate the cumulative probabilities, assuming that all of
the upper limits have $\log R_{8.4}$ values less than the lowest
detection.

\pagebreak

%
%

\begin{table}
\caption{\label{tab:tablerad} }
\end{table}

\begin{table}
\caption{\label{tab:tableopt} }
\end{table}

\begin{figure}
\caption{\label{fig:zMBLBQS} } 
\end{figure}

\begin{figure}
\caption{\label{fig:LnR_MB} } 
\end{figure}
 
\begin{figure}
\caption{\label{fig:smoothMB} } 
\end{figure}

\begin{figure}
\caption{\label{fig:zMB} }  
\end{figure}

\begin{figure}
\caption{\label{fig:smoothzLBQS} } 
\end{figure}

\begin{figure}
\caption{\label{fig:smoothzEMSS} } 
\end{figure}

\begin{figure}
\caption{\label{fig:LBQScum} } 
\end{figure}


\begin{references}

\reference{SkiARAA} Antonucci, R. 1993, \araa, 31, 473

\reference{Bahcall96} Bahcall, J. N., Kirhakos, S., \& Schneider, D.
P. 1996, \apj, 457, 557

\reference{Barthel89} Barthel, P. D. 1989, \apj, 336, 606

\reference{Blandford77} Blandford, R. D., \& Znajek, R. L. 1977,
\mnras, 179, 433

\reference{Brinkmann95} Brinkmann, W., Siebert, J., Reich, W.,
F\"urst, E., Reich, P., Voges, W., Tr\"umper, J., Wielebinski, R.
1995, \aaps, 109, 147

\reference{Coleman92} Coleman, C. S., \& Dopita, M. A. 1992, Proc.
Astron. Soc. Australia, 10, 107

\reference{Condon81} Condon, J. J., O'Dell, S. L., Puschell, J. J., \&
Stein, W. A. 1981, \apj, 246, 624

\reference{Condon89} Condon, J. J., Broderick, J. J., Seielstad, G.
A. 1989, \aj, 97, 1064

\reference{Della94} Della Ceca, R., Zamorani, G., Maccacaro, T., Wolter, A.,
Griffiths, R., Stocke, J. T., \& Setti, G. 1994, \apj, 430, 533

\reference{Disney95} Disney, M. J., et al. 1995, \nat, 376, 150

\reference{Dunlop90} Dunlop, J. S., \& Peacock, J. A. 1990, \mnras,
247, 19

\reference{Falcke95} Falcke, H., Gopal-Krishna, Biermann, P. L. 1995, 
\aap, 298, 395

\reference{Francis93} Francis, P. J. 1993, \apj, 407, 519

\reference{Francisetal91} Francis, P. J., Hewett, P. C., Foltz, C. B., Chaffee, F.
H., Weymann, R. J., \& Morris, S. L. 1991, \apj, 373, 465

\reference{Gioia90} Gioia, I. M., Maccacaro, T., Schild, R. E.,
Wolter, A., Stocke, J. T., Morris, S. L., \& Henry, J. P. 1990,
\apjs, 72, 567

\reference{Hewett95} Hewett, P. C., Foltz, C. B., \& Chaffee, F. H. 1995,
\aj, 109, 1498

\reference{PaperII} Hooper, E. J., Impey, C. D., Foltz, C. B., \& Hewett, P. C.
1995, \apj, 445, 62 (Paper II)

\reference{Hutchings94} Hutchings, J. B., Holtzman, J., Sparks, W. B.,
Morris, S. C., Hanisch, R. J., \& Mo, J. 1994, \apj, 429, L1

\reference{Kellermann89} Kellermann, K. I., Sramek, R., Schmidt, M.,
Shaffer, D. B., \& Green, R. 1989, \aj, 98, 1195 (PG)

\reference{LaF94} La Franca, F., Gregorini, L., Cristiani, S., 
De Ruiter, H., \& Owen, F. 1994, \aj, 108, 1548

\reference{Marshall87} Marshall, H. L. 1987, \apj, 316, 84

\reference{McMahon92} McMahon, R. G., Irwin, M. J., \& Hazard, C.
1992, in X-ray Emission from Active Galactic Nuclei and the Cosmic
X-ray Background, ed. W. Brinkmann \& J. Tr\"umper (Garching:
Max-Planck-Institut f\"ur extraterrestrische Physik), 399

\reference{MPM} Miller, L., Peacock, J. A., \& Mead, A. R. G. 1990,
\mnras, 244, 207

\reference{Padovani93} Padovani, P. 1993, \mnras, 263, 461

\reference{Peacock86} Peacock, J. A., Miller, L., \& Longair, M. S.
1986, \mnras, 218, 265

\reference{Rees84} Rees, M. J. 1984, \araa, 22, 471

\reference{Sandage65} Sandage, A. 1965, \apj, 141, 1560

\reference{Schmidt95} Schmidt, M., van Gorkom, J. H., Schneider, D.
P., \& Gunn, J. E. 1995, \aj, 109, 473

\reference{Schneider92} Schneider, D. P., van Gorkom, J. H., Schmidt,
M., \& Gunn, J. E. 1992, \aj, 103, 1451

\reference{Smith86} Smith, E. P., Heckman, T. M., Bothun, G. D.,
Romanishin, W., \& Balick, B. 1986, \apj, 306, 64

\reference{SW80} Sramek, R. A., \& Weedman, D. W. 1980, \apj, 238, 435 

\reference{Stetson87} Stetson, P. B. 1987, \pasp, 99, 191

\reference{Stetson90} Stetson, P. B. 1990, \pasp, 102, 932

\reference{Stocke91} Stocke, J. T., Morris, S. L., Gioia, I. M.,
Maccacaro, T., Schild, R., Wolter, A., Fleming, T. A., \& Henry, J. P.
1991, \apjs, 76, 813

\reference{Stocke92} Stocke, J. T., Morris, S. L., Weymann, R. J., 
\& Foltz, C. B. 1992, \apj, 396, 487

\reference{Urry91} Urry, C. M., Padovani, P., \& Stickel, M. 1991,
\apj, 382, 501

\reference{VCW90} V\'{e}ron-Cetty, M. P., \& Woltjer, L. 1990, \aap, 236, 69 

\reference{Visnovsky92} Visnovsky, K. L., Impey, C. D., Foltz, C. B., 
Hewett, P. C., Weymann, R. J., \& Morris, S. L. 
1992, \apj, 391, 560 (Paper I)

\reference{Voges92} Voges, W. 1992, in Proceedings of the European
International Space Year Conference ``Space sciences with particular
emphasis on high-energy astrophysics,'' ESA ISY-3, ed. T. D. Guyenne
\& J. J. Hunt (Paris: ESA Publications), 9  
  
\reference{Wilson95} Wilson, A. S., \& Colbert, E. J. M. 1995, \apj,
438, 62

\reference{Worrall87} Worrall, D. M., Giommi, P., Tananbaum, H., \&
Zamorani, G. 1987, \apj, 313, 596
 
\end{references}
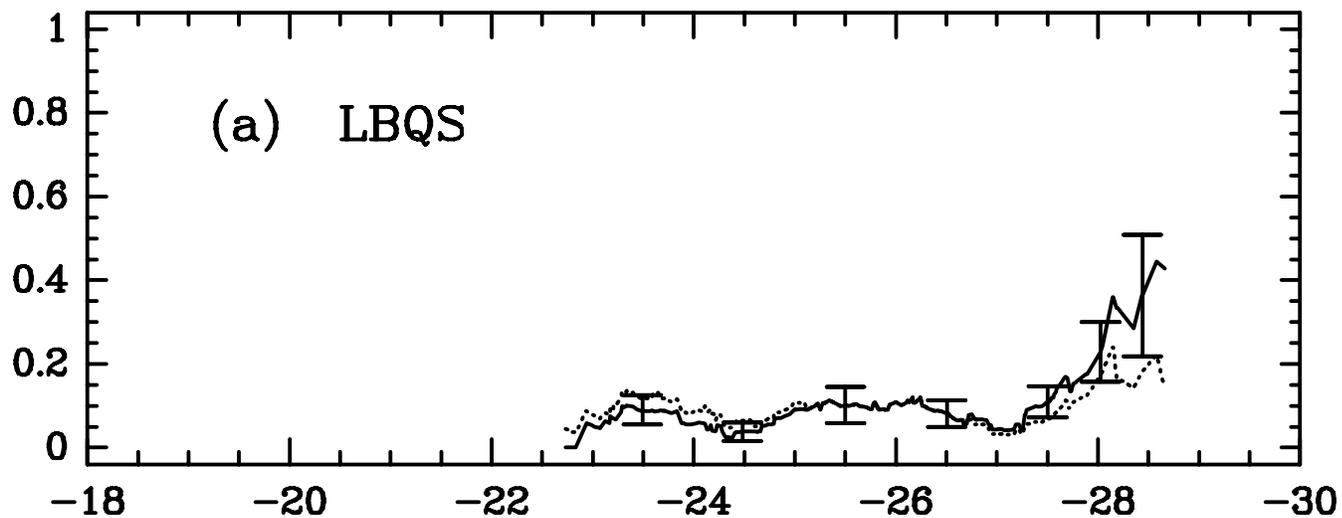
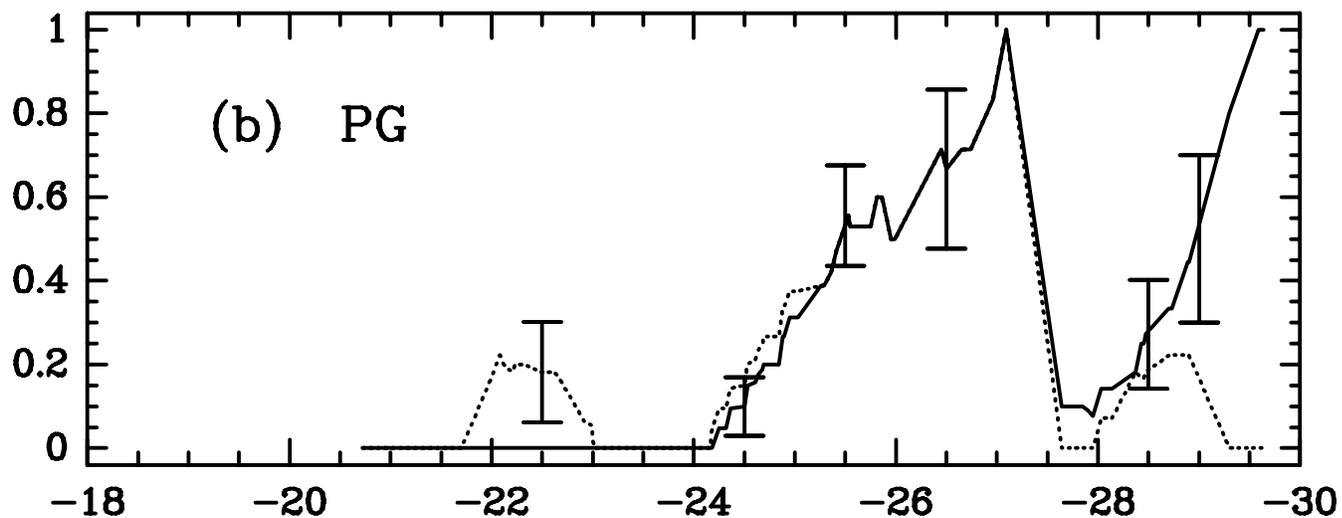
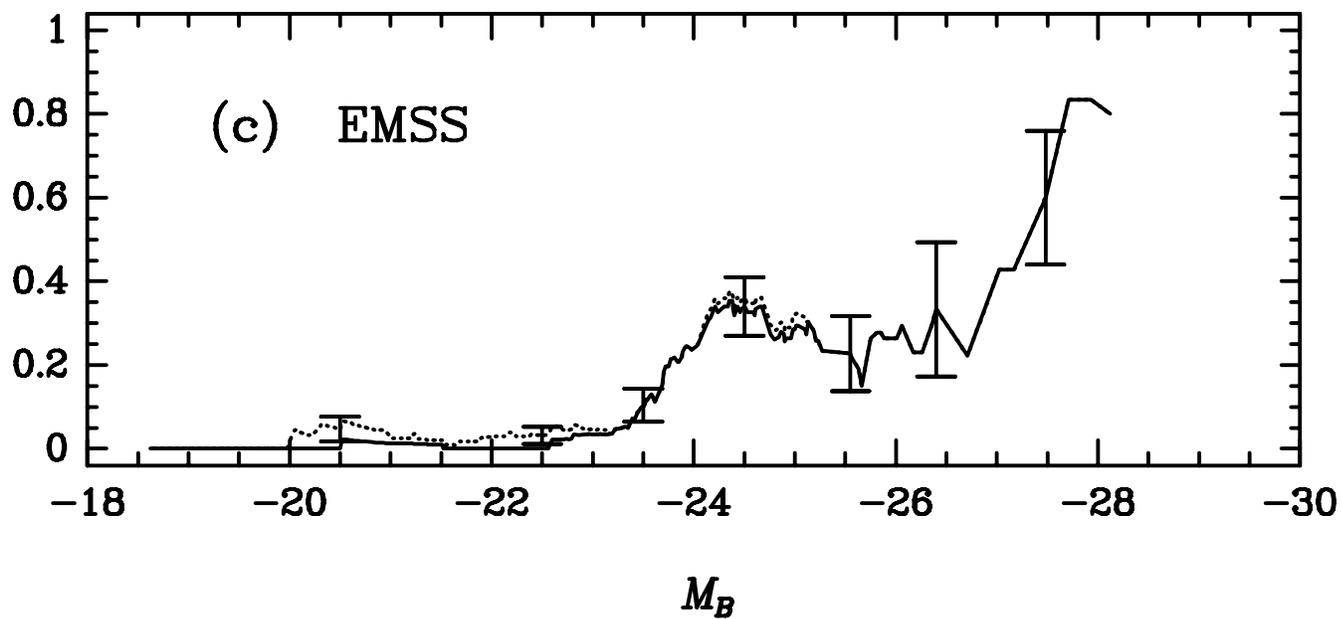
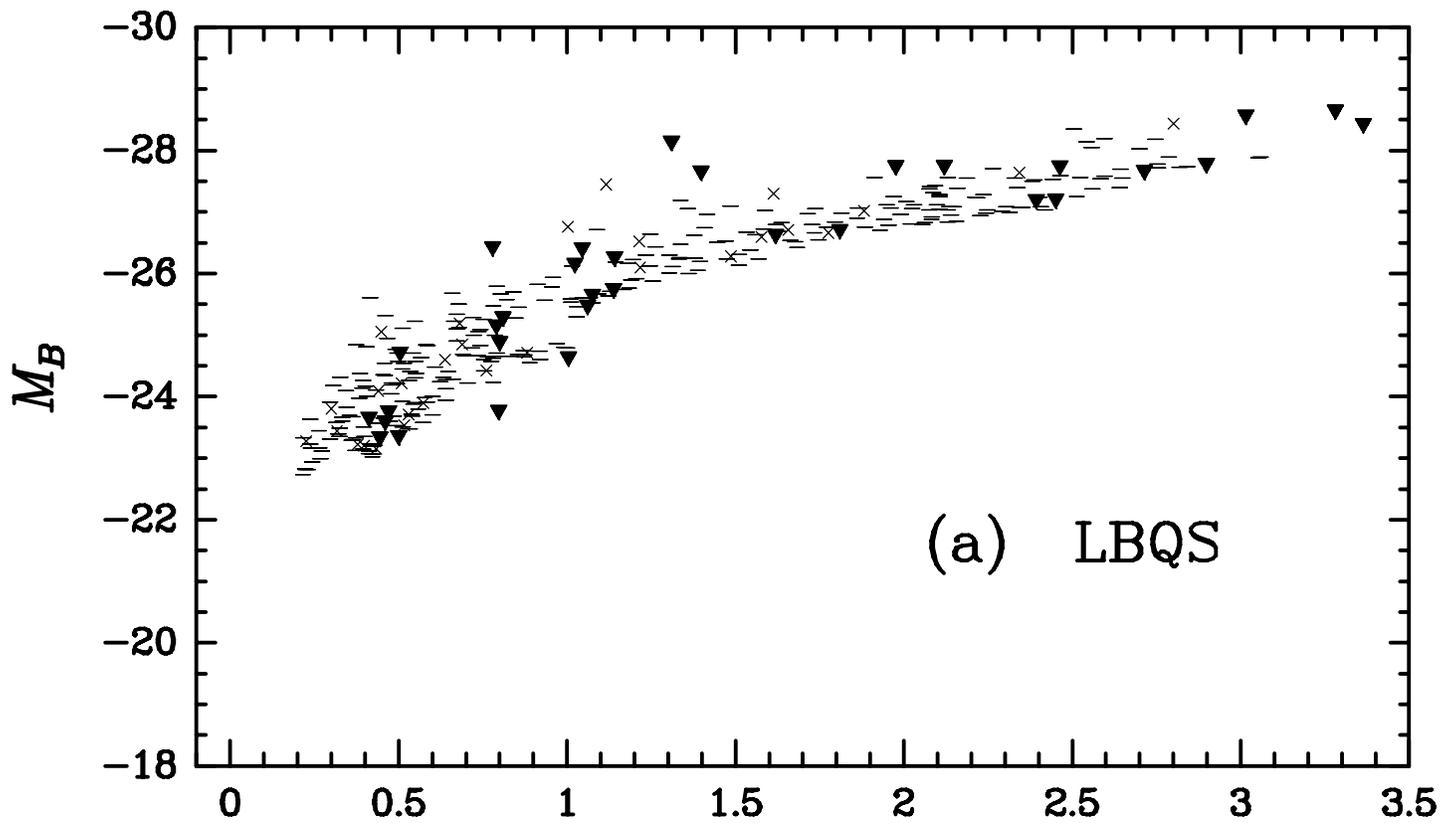
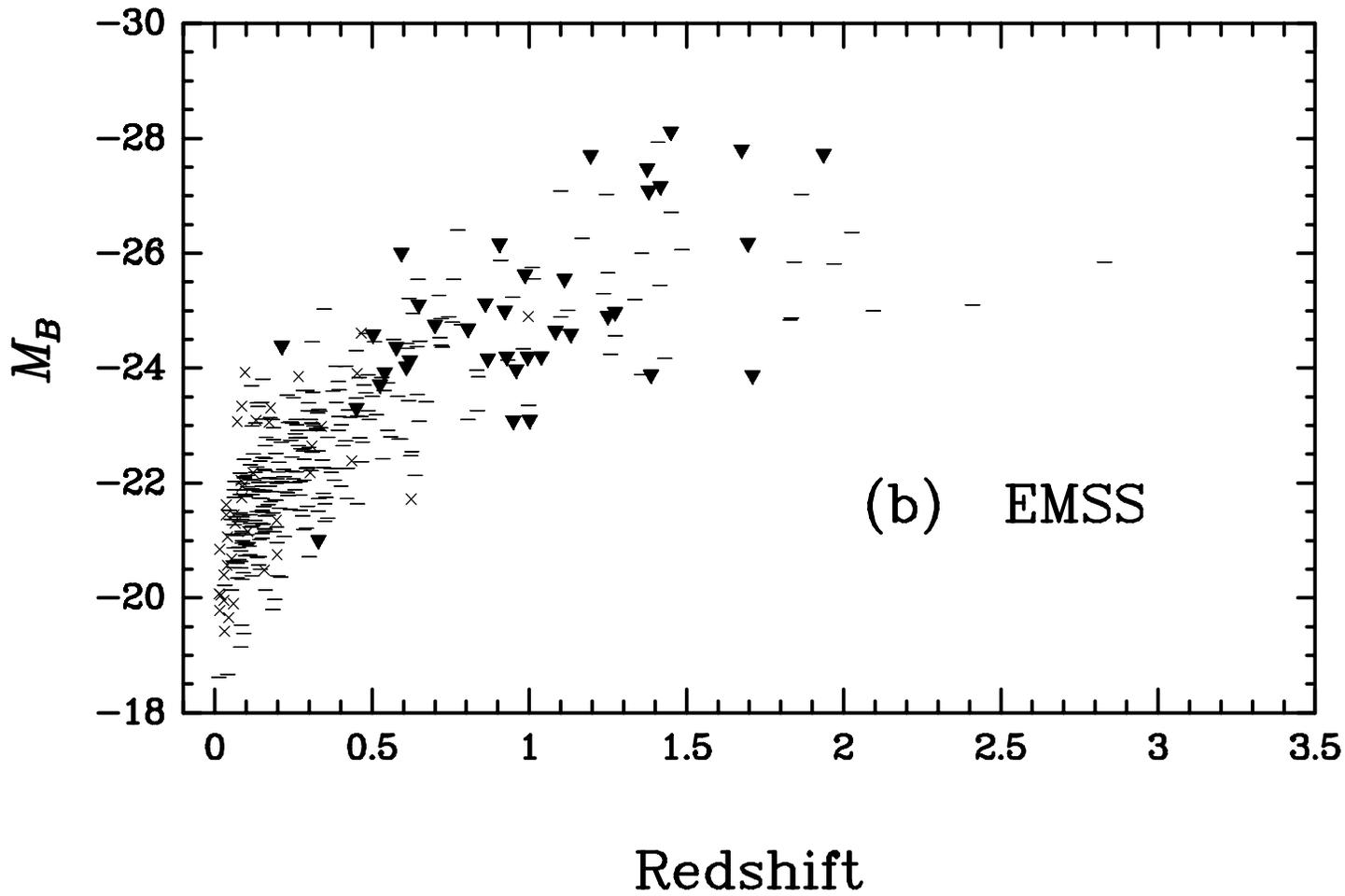
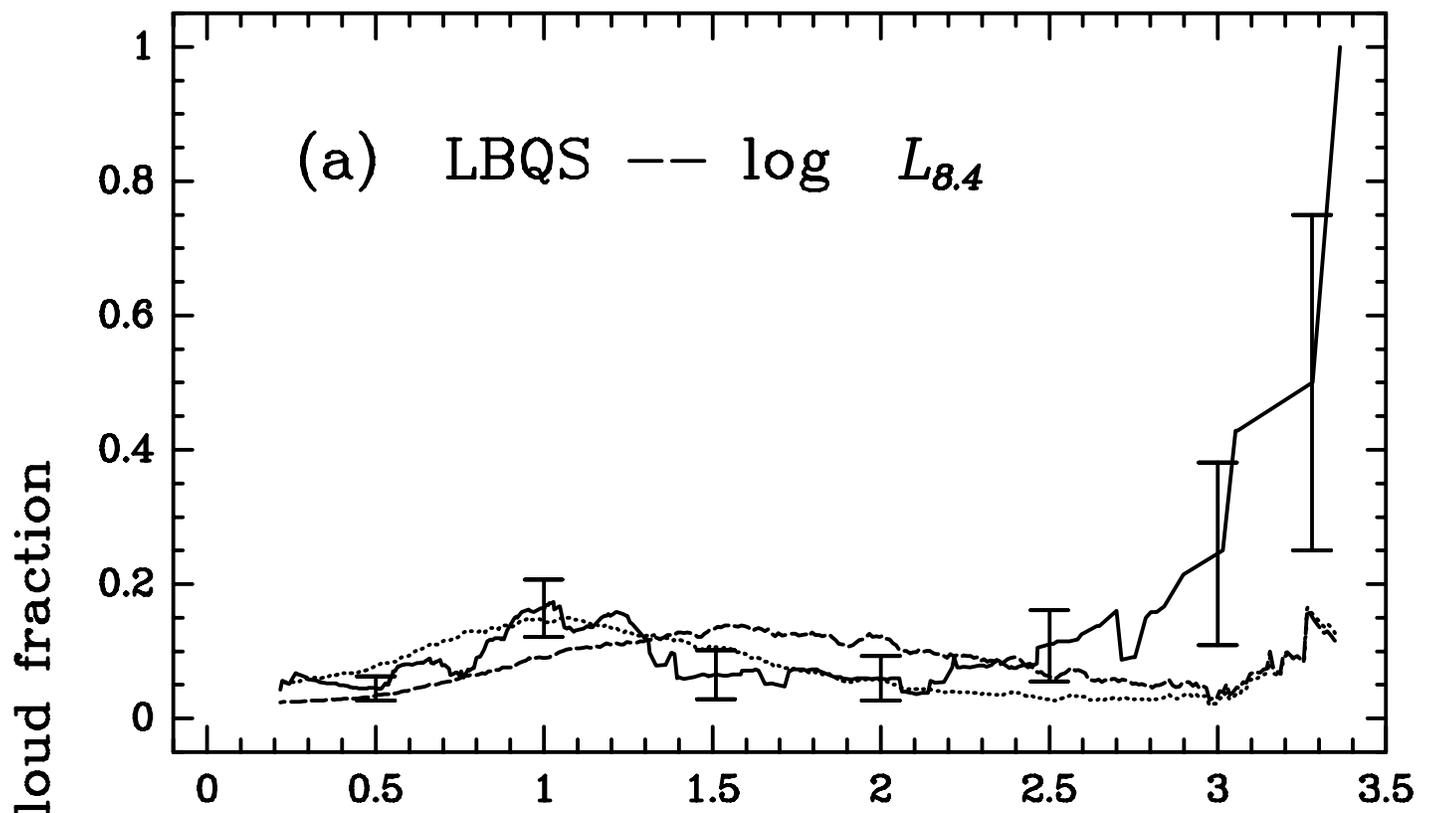
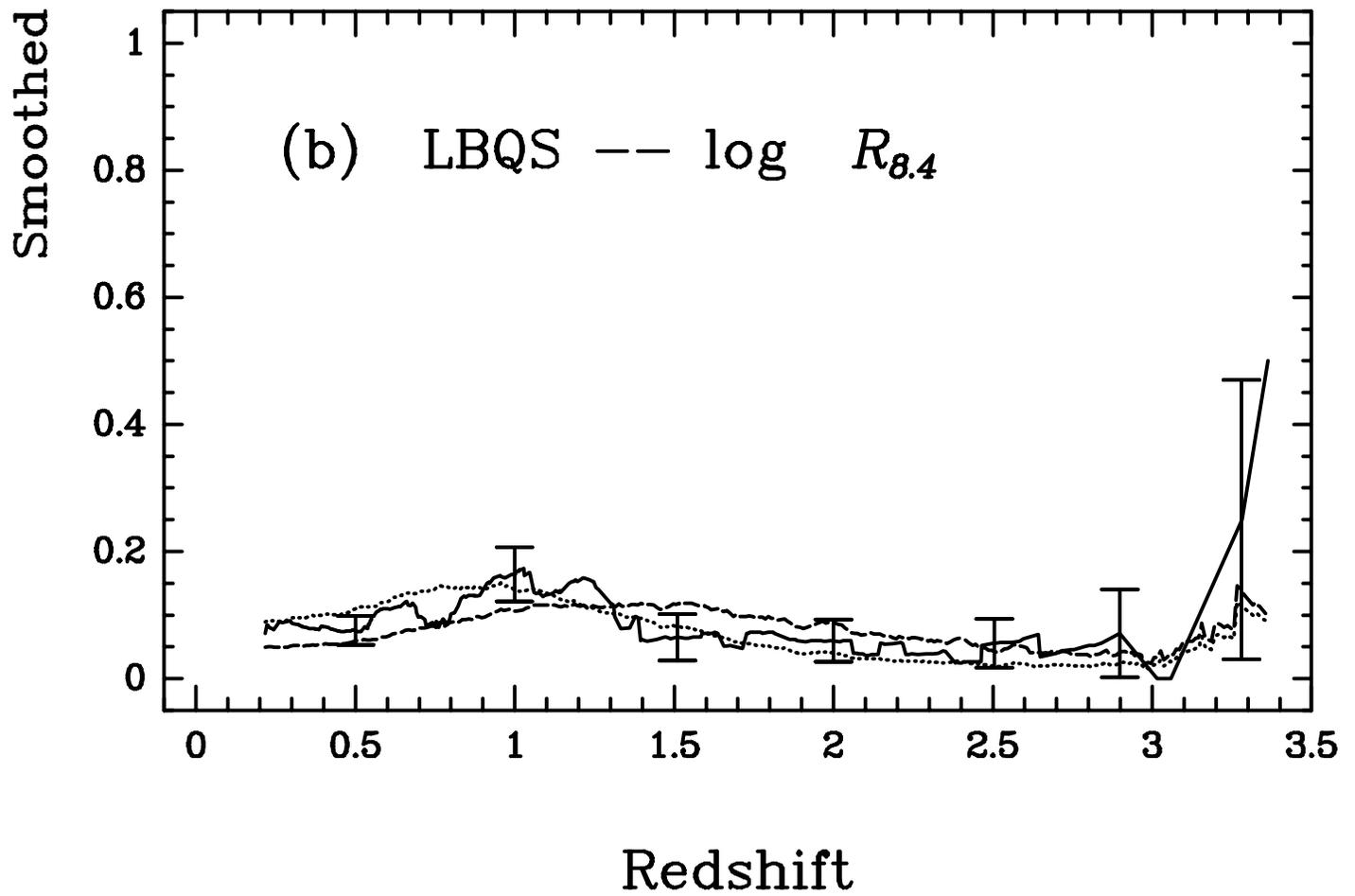
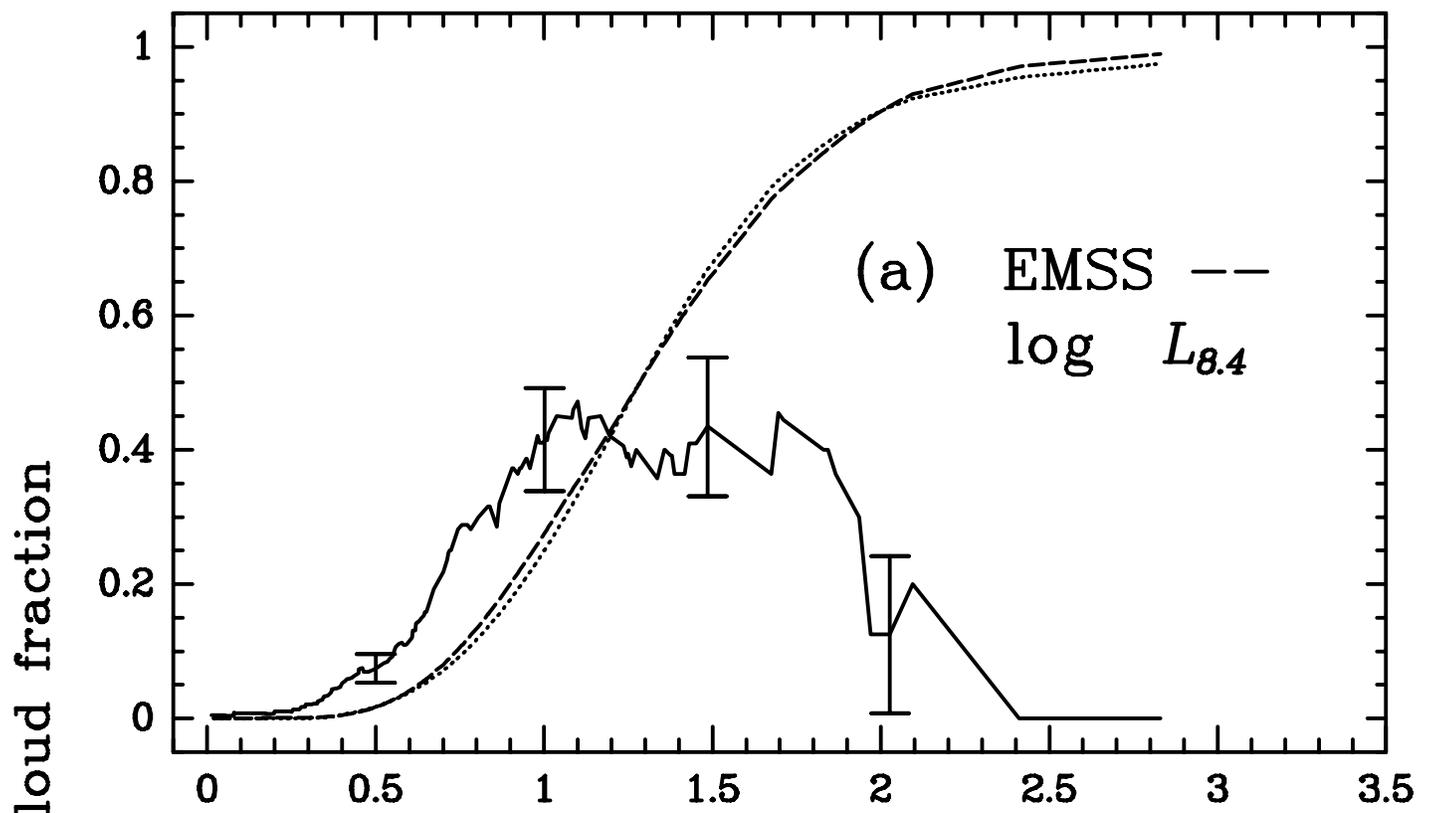
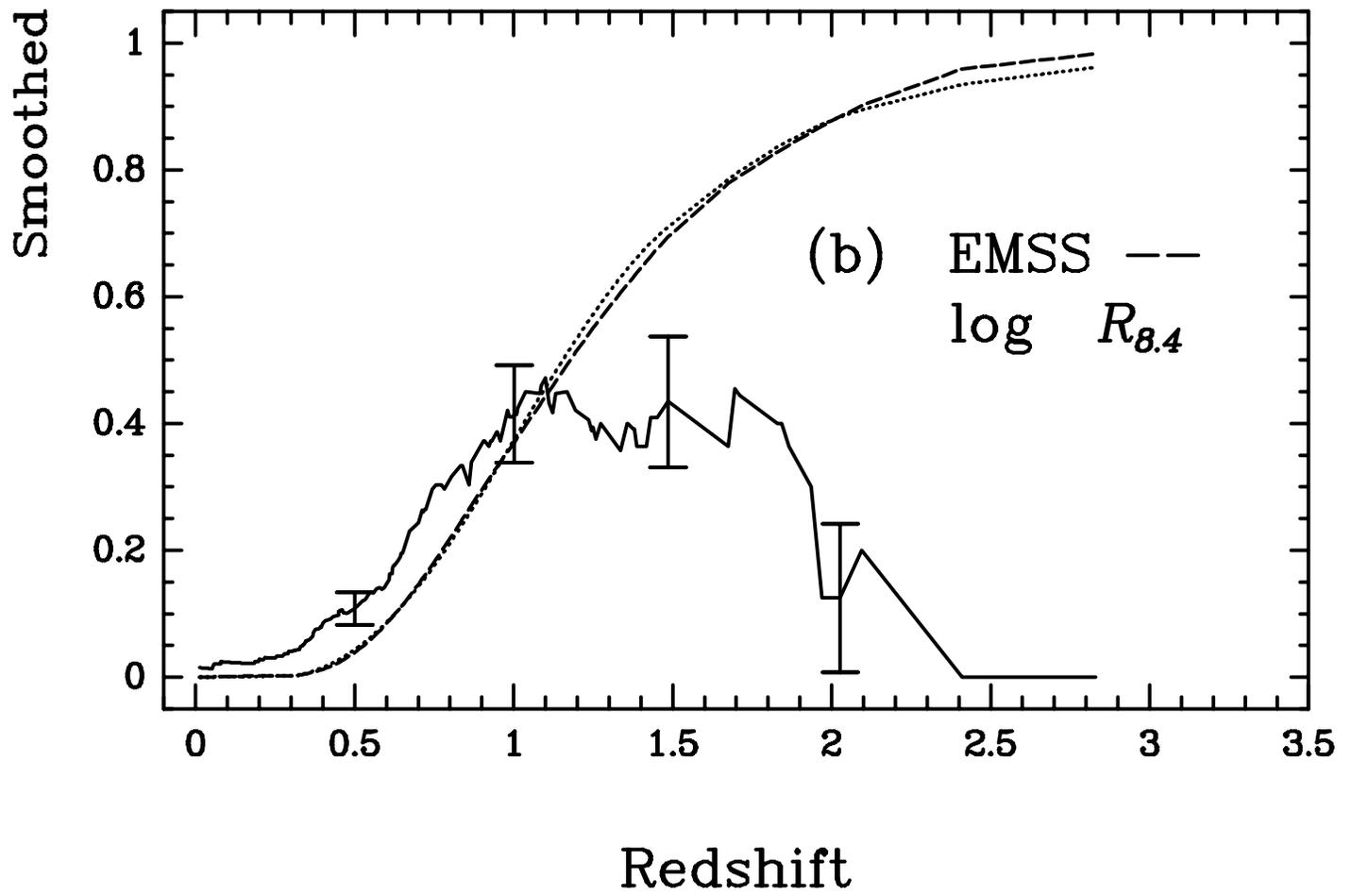
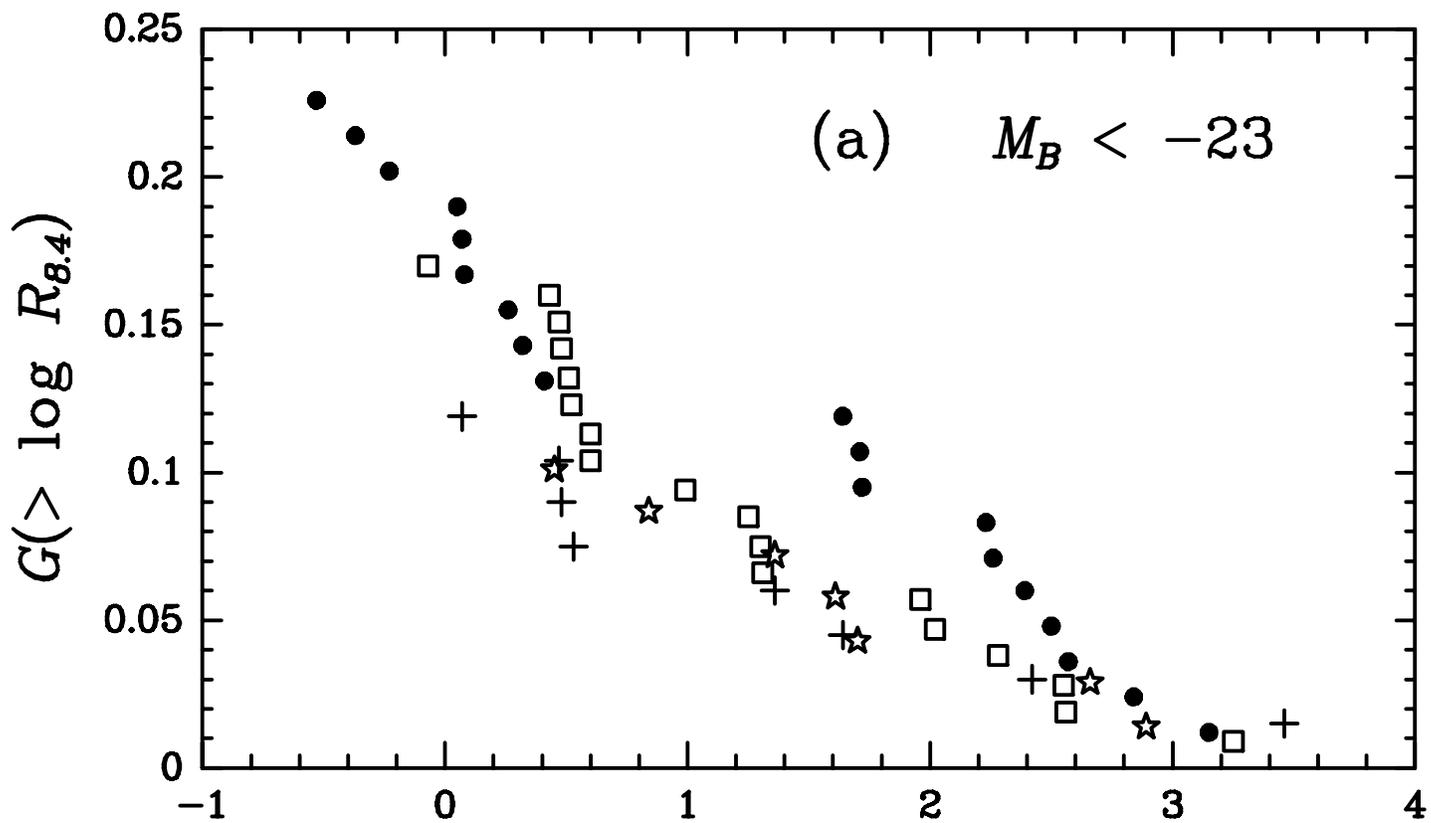
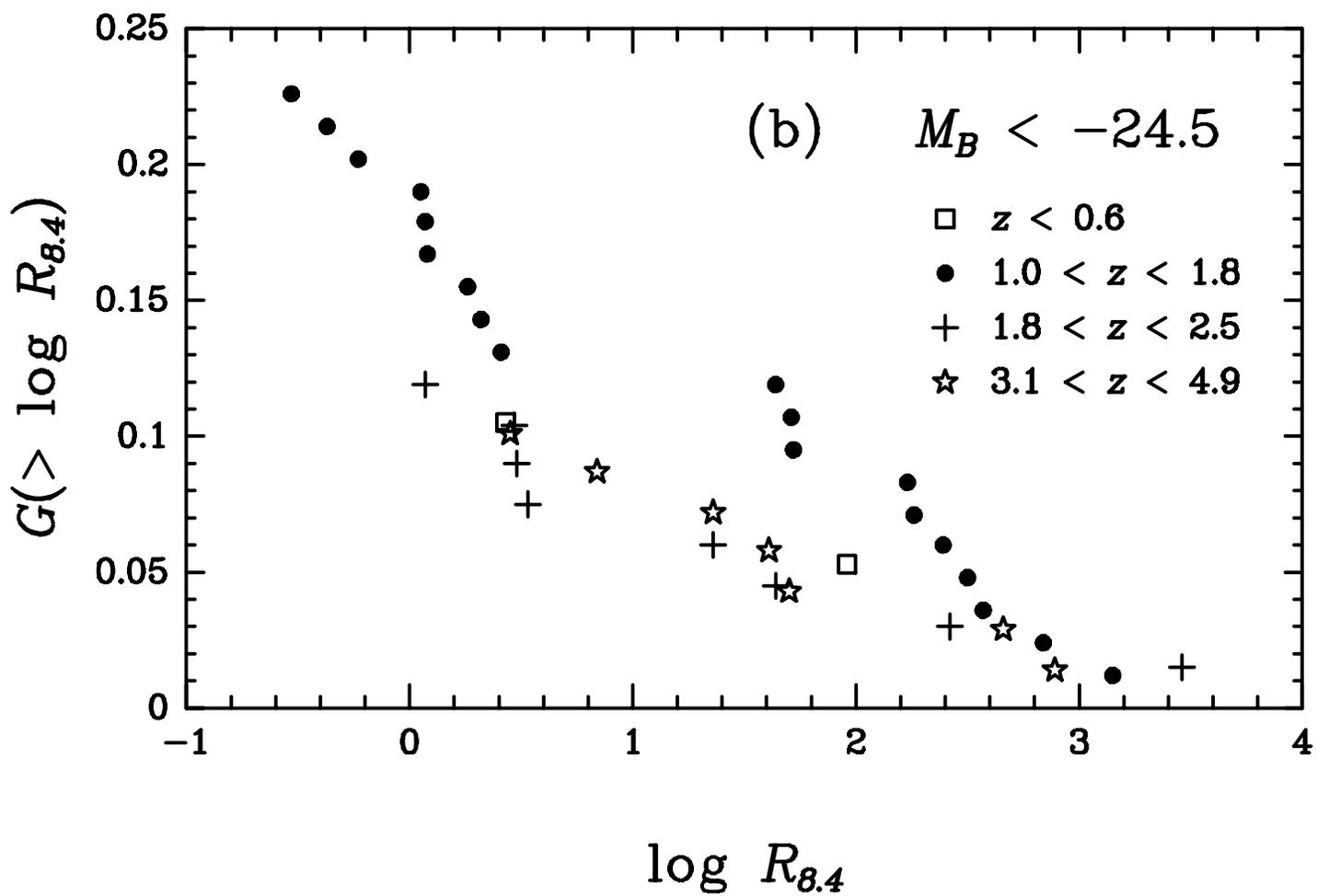
\end{document}